\documentstyle[11pt]{article}

\begin{document}

\input{epsf}

\def\beq{\begin{equation}}
\def\eeq{\end{equation}}
\def\bea{\begin{eqnarray}}
\def\eea{\end{eqnarray}}
\def\beas{\begin{eqnarray*}}
\def\eeas{\end{eqnarray*}}
\def\ov{\overline}
\def\ot{\otimes}

\newcommand{\hf}{\mbox{$\frac{1}{2}$}}
\def\sig{\sigma}
\def\De{\Delta}
\def\af{\alpha}
\def\be{\beta}
\def\la{\lambda}
\def\ga{\gamma}
\def\ep{\epsilon}
\def\vep{\varepsilon}
\def\half{\frac{1}{2}}
\def\third{\frac{1}{3}}
\def\fth{\frac{1}{4}}
\def\sth{\frac{1}{6}}
\def\tth{\frac{1}{24}}
\def\tde{\frac{3}{2}}

\def\nbar{{\bar{n}}}
\def\zb{{\bar z}} 
\def\psib{{\bar \psi}} 
\def\etab{{\bar \eta }}
\def\gab{{\bar \ga}}
\def\vev#1{\langle #1 \rangle}
\def\inv#1{{1 \over #1}}
\def\ots{{\stackrel{s}{\otimes}}}

\def\CA{{\cal A}}       \def\CB{{\cal B}}       \def\CC{{\cal C}}
\def\CD{{\cal D}}       \def\CE{{\cal E}}       \def\CF{{\cal F}}
\def\CG{{\cal G}}       \def\CH{{\cal H}}       \def\CI{{\cal J}}
\def\CJ{{\cal J}}       \def\CK{{\cal K}}       \def\CL{{\cal L}}
\def\CM{{\cal M}}       \def\CN{{\cal N}}       \def\CO{{\cal O}}
\def\CP{{\cal P}}       \def\CQ{{\cal Q}}       \def\CR{{\cal R}}
\def\CS{{\cal S}}       \def\CT{{\cal T}}       \def\CU{{\cal U}}
\def\CV{{\cal V}}       \def\CW{{\cal W}}       \def\CX{{\cal X}}
\def\CY{{\cal Y}}       \def\CZ{{\cal Z}}

\newcommand{\np}{Nucl. Phys.}
\newcommand{\pl}{Phys. Lett.}
\newcommand{\prl}{Phys. Rev. Lett.}
\newcommand{\cmp}{Commun. Math. Phys.}
\newcommand{\jmp}{J. Math. Phys.}
\newcommand{\jpamg}{J. Phys. {\bf A}: Math. Gen.}
\newcommand{\lmp}{Lett. Math. Phys.}
\newcommand{\ptp}{Prog. Theor. Phys.}

\newif\ifbbB\bbBfalse                
\bbBtrue                             

\ifbbB   
 \message{If you do not have msbm (blackboard bold) fonts,}
 \message{change the option at the top of the text file.}
 \font\blackboard=msbm10 
 \font\blackboards=msbm7 \font\blackboardss=msbm5
 \newfam\black \textfont\black=\blackboard
 \scriptfont\black=\blackboards \scriptscriptfont\black=\blackboardss
 \def\Bbb#1{{\fam\black\relax#1}}
\else
 \def\Bbb{\bf}
\fi

\def\id{{1\! \! 1 }}
\def\bo{{\Bbb 1}}
\def\bI{{\Bbb I}}
\def\bC{{\Bbb C}} 
\def\bZ{{\Bbb Z}}
\def\CN{{\cal N}}

\title{Fermionization and Hubbard Models}
\author{{\bf P. Dargis} and
{\bf Z. Maassarani}\thanks{Email address: zmaassar@phy.ulaval.ca}\\
\\
{\small D\'epartement de Physique, Pav. A-Vachon}\\
{\small Universit\'e Laval,  Ste Foy, Qc,  
G1K 7P4 Canada}\thanks{Work supported by NSERC 
(Canada) and FCAR (Qu\'ebec).} \\}
\date{}
\maketitle

\begin{abstract}
We introduce a  transformation which allows
the fermionization of operators of 
any one-dimensional spin-chain. This fermionization
procedure  is independent of any eventual integrable structure and  is
compatible with it. We  illustrate this method on various
integrable and non-integrable chains, and deduce some   
general results.     
In particular, we  fermionize XXC spin-chains and study their symmetries.
Fermionic realizations of certain Lie algebras and 
superalgebras appear naturally as symmetries
of some models.  
We also fermionize recently obtained Hubbard models, and obtain for the first 
time multispecies analogues of the Hubbard model, in their fermionic form.
We comment on the conflict between  symmetry enhancement 
and integrability of these models.
Finally, the fermionic versions of the  non integrable 
spin-$1$ and spin-$\frac{3}{2}$ Heisenberg chains are  obtained.

\end{abstract}
\vspace*{2.5cm}
\noindent
\hspace{1cm} May 1998\hfill\\
\hspace*{1cm} LAVAL-PHY-21/98\hfill\\
\hspace*{1cm} cond-mat/9806208

\vspace*{2.5cm}
\noindent
\hspace*{1cm} PACS numbers: 02.90.+p, 05.50.+q, 74.20.-z, 71.10.Fd\hfill\\
\hspace*{1cm} Key words: Fermionization, Spin-chains, 
Integrability, Hubbard models\hfill\\

\thispagestyle{empty}

\newpage

\setcounter{page}{1}

\section{Introduction}

Fermionization and its reverse procedure, bosonization,
consist of a mapping between a set of bosonic variables and one of
fermionic variables. Such mappings  have long
been  essential tools of theoretical physics. They provide 
a way of expressing the same theory in two different languages
and open the door to the use of a large number 
of techniques to study a particular system. 
They also reveal the interplay  between symmetries and boundary conditions,
and provide new models, in particular when the dimension of space or
the underlying lattice is changed.
Fermionization schemes exist in many dimensions. They
take an added importance in  two-dimensional conformal field
theories; see for instance \cite{fms}.

For the  spin-$\frac{1}{2}$ one-dimensional quantum spin-chain  
a fermionization 
procedure exists which allows the mapping between spin operators and
fermionic creation-annihilation operators. 
This  Jordan-Wigner transformation \cite{jowi} has been
used in particular to study the polaron model \cite{puzh}  and
the Hubbard model \cite{sh12}. 
The fermionic forms appear as models of itinerant 
electrons used to model conduction properties,  
while the bosonic ones appear as spin-chains.
More recently the Jordan-Wigner transformation 
was used in the study of the  Bariev model 
of correlated-hopping electrons \cite{bariev}. 
However the Jordan-Wigner transformation is limited in scope and
has long been standing alone 
as a way to fermionize simple one-dimensional spin-chains.
It is one of  the aims of this paper to provide a general fermionization 
method. 

We start by recalling the  Jordan-Wigner transformation and
then give its generalization. We show how several variants  can be
implemented to fermionize operators corresponding
to any spin-chain, integrable or not. 
We then review how to fermionize, with the help of the Jordan-Wigner
mapping,  an integrable spin-chain model
of considerable  theoretical and experimental interest,
the spin-$\frac{1}{2}$ Heisenberg spin-chain Hamiltonian.
As boundary conditions are not preserved by the Jordan-Wigner transformation,
we  clarify this issue  on this simple example. 
The fermionic version  possesses non-local boundary terms. 
It is possible to replace these terms with periodic ones without spoiling 
integrability. This example serves also other
purposes. The general features of integrability, both in the
bosonic and fermionic settings,  are introduced and  
quantities and equations which will reappear  for other models are
defined for this simple model.
We  also  obtain some general  results by
studying this system.
We then fermionize three of the recently found
integrable XXC models and the $su(4)$ XXZ model, {\it i.e.\/} the
model corresponding to the $R$-matrix of the fundamental representation
of $su(4)$. The XXC models are  hybrid trigonometric ones,
with an underlying $su(2)$ structure and some $su(n)$ features. 
They possess  symmetries much larger
than the trigonometric models built
from the fundamental representations of $su(n)$. 
We apply the fermionization at the Hamiltonian level and 
at the integrable structure level, by fermionizing the $L$- and $R$-matrices.
While different versions of a given XXC model exist, they
are known to be equivalent at the original bosonic level.  This
turns out to be false  for most  fermionized versions. 
We also show that fermionic realizations of algebras and  
superalgebras  appear naturally as the symmetries of the  fermionized 
models. Some symmetries are  broken only to be replaced by new ones. 
From the examples studied we infer some  general results. 

The XXC models at their `free-point' are known to be the building blocks
of multistates generalizations of the Hubbard model \cite{zm12,ff}.
These generalizations inherit the large symmetries of the component 
models. 
This is another motivation to study fermionized versions of the XXC models.  
We recall the fermionization of the usual Hubbard model and
do the same for two   generalized Hubbard Hamiltonians and
their $L,R$-matrices. This provides for  the first time fermionic 
multiple-species generalizations of the Hubbard models. 
The Hamiltonian density can then be written on a lattice of any dimension
and provides possible candidates for superconducting  models in two dimensions. 

The usual Hubbard model has a $U(1)\times  U(1)$ symmetry in both 
its bosonic and fermionic forms. 
In its original fermionic definition, this symmetry is enhanced to
an $SO(4)$  group symmetry.  
We  investigate whether a similar  enhancement 
takes place for our generalized Hubbard  models. 
 
We then derive the fermionic versions of two models of considerable
physical importance, the spin-$1$ and spin-$\frac{3}{2}$ XXZ 
Heisenberg chains. We  propose for the Hamiltonians slightly modified versions
which do not contain non-local factors.

We conclude with general considerations and remarks 
about  the foregoing fermionization scheme. 
Finally we  provide a preliminary comparison with 
a, different,  recently introduced fermionization scheme for integrable
models \cite{gomu}.

\section{Fermionization}\label{newmod}

The mapping from purely bosonic operators to fermionic ones
may seem to require  the Hilbert space, at a given site, to
have as dimension a power of two. This is obvious 
on general grounds. The Fock space of a set of anticommuting 
Fermi operators   has dimension $2^d$ where $d$ is the number
of species of spinless fermions, {\it i.e.} the number of pairs of mutually
anticommuting creation-annihilation operators. 
However, a local space with  arbitrary dimension can be embedded in the nearest
larger space of dimension a power of two. The fermionization   is then
done in the latter space, with an eventual projection on 
the original space. We shall illustrate this procedure on various 
examples.  
Before going any farther we recall the
Jordan-Wigner transformation for a chain of spin-$\frac{1}{2}$ 
variables \cite{jowi}.

\subsection{Jordan-Wigner transformation}

Let $\sig^{\pm}$ and $\sig^z$ be the three Pauli matrices:
\beq
\sig^+=       \left(\begin{array}{cc}
            0&1\\
            0&0
       \end{array} \right)
\;\;\;\;\;\;\sig^-=       \left(\begin{array}{cc}
            0&0\\
            1&0
       \end{array} \right)
\;\;\;\;\;\;\sig^z=       \left(\begin{array}{cc}
            1&0\\
            0&-1
       \end{array} \right)
\eeq
The transformation at the site $m$ is non-local and given by 
\beq
a_m^\dagger = \pi^{\sig}_{m-1}\sig^+_m \quad ,\quad
a_m = \pi^{\sig}_{m-1}\sig^-_m
\eeq
with  
\beq
\pi^{\sig}_0\equiv \bI\quad\quad {\rm and}\quad 
\quad \pi^{\sig}_{m-1}\equiv\sig^z_1 \cdots \sig_{m-1}^z
\eeq 
where $\bI$ is the identity operator. 
It is straightforward to check that the new operators are fermionic
creation-annihilation operators,
\beq
\{a_m,a_n\} =\{a_m^\dagger,a_n^\dagger \}=0 \quad ,\quad 
\{a_m^\dagger,a_n \} = \delta_{mn}
\eeq
where $\{\;,\;\}$ denotes the anticommutator, $\{A,B\}=A\, B+B\, A$.
One may also have more than a single copy of the Pauli matrices
at every site. For instance, one may have two copies at every site resulting
in a local space  of dimension 4. 
This is what happens when one bosonizes the one-dimensional Hubbard model. 
The spin `up' and `down' fermions  result in two copies 
of the Pauli matrices. The $p$-copies  Bariev model contains
$p$ copies of the spin-$\frac{1}{2}$ representation at every site \cite{bariev}.
Its fermionization results in $p$ species of spinless fermions.
The fermionization formulae for an $N$-sites chain  are given by:
\beq
\begin{array}{ll}
a_{1m}^\dagger = \pi^{(1)}_{m-1}\sig^{(1)+}_m \;\;,\;\; 
&a_{1m} = \pi^{(1)}_{m-1}\sig^{(1)-}_m \\
a_{2m}^\dagger = \pi^{(1)}_{N+1}\pi^{(2)}_{m-1}\sig^{(2)+}_m \;\;,\;\; 
&a_{2m} = \pi^{(1)}_{N+1}\pi^{(2)}_{m-1}\sig^{(2)-}_m \\
...................................
&................................... \\
a_{pm}^\dagger = \pi^{(1)}_{N+1}\cdots \pi^{(p-1)}_{N+1}\pi^{(p)}_{m-1}
\sig^{(p)+}_m \;\;,\;\; &a_{pm} = \pi^{(1)}_{N+1}\cdots \pi^{(p-1)}_{N+1}\pi^{(p)}_{m-1} \sig^{(p)-}_m 
\end{array}\label{multiple}
\eeq
where $\pi^{(s)}_{m-1}\equiv\sig^{(s)z}_1 \cdots \sig_{m-1}^{(s)z}$,
$s=1,...,p$. The $\pi^{(s)}_{N+1}$ ensure anticommutation between different
species:
\beq
\{a_{sm},a_{tn} \} =\{a_{sm}^\dagger,a_{tn}^\dagger \}=0 
\quad,\quad \{a_{sm}^\dagger,a_{tn} \} = \delta_{mn}\delta_{st}
\eeq

The Pauli matrices satisfy the $su(2)$ algebra. However, $\sig^x$ 
and $\sig^y$ also satisfy the Clifford algebra and
this is the relevant remark which allowed us to generalize the
Jordan-Wigner transformation.

\subsection{Generalization}

Let $d$ be a positive integer. 
The dimension $2d$\/ Clifford algebra, ${\cal C}_{2d}$, is generated by 
an even number of  
mutually anticommuting generators
\beq
\{ \ga_i,\ga_j\}= +2\,\delta_{ij}\;\;\;,\;\;\; i,j=1,...,2d
\eeq
It is a classical exercise to show that this algebra contains
the following $2^{2d}$ linearly independent elements:
\beq
\bI\; ; \;\; \gamma_i\; ;\;\; 
\ga_{i_1}\ga_{i_2}\;,\;\; i_1<i_2\; ;\;\; \cdots\cdots\; ; 
\ga_{i_1}\!\cdots\ga_{i_{2d}}\;,\;\; i_1<\cdots <i_{2d}
\eeq
The $2d$ generators of this algebra can be realized as
$2^d\times 2^d$ hermitian matrices. Thus {\it all \/}   
$2^d\times 2^d$ matrices
can be written in a unique way as linear combinations
of  products of $\ga$ matrices.
For $d=1$ we recover the Pauli matrices while for $d=2$ we get
the Dirac $\gamma$-matrices.
There is a special element of ${\cal C}_{2d}$ which anticommutes 
with all  elements $\ga^i$, and squares to one. It is the product
of the $2d$\/ $\ga$-matrices; we take a particular normalization for
convenience:
\beq
\ga\equiv i^{-d}\,\ga_1\cdots \gamma_{2d}
\eeq
Let  
\beq
\ga^{\pm}_{j}\equiv \frac{1}{2} (\ga_{2j-1}\pm i\ga_{2j})\;\;,\;\;\;j=1,...,d
\eeq
We define a generalized  Jordan-Wigner transformation
by the following mapping between bosonic matrices and 
fermionic creation-annihilation operators  of $d$ species
of spinless fermions:
\beq
a_{jm}^\dagger =\pi^{\ga}_{m-1}\ga^+_{jm} \;\;,\;\; a_{jm} = \pi^{\ga}_{m-1}\ga^-_{jm}\;\;,\;\;\;j=1,...,d\label{dm}
\eeq
where 
\beq
\pi^{\ga}_0\equiv \bI \quad\quad {\rm and}\quad\quad
\pi^{\ga}_{m-1}=\ga_1\cdots \ga_{m-1}\label{dm1}
\eeq
It is straightforward to verify that the following anticommutation
relations hold:
\beq
\{a_{jm},a_{kn} \} =\{a_{jm}^\dagger,a_{kn}^\dagger \}=0 \;\;,\;\;\ 
\{a_{jm}^\dagger,a_{kn} \} = \delta_{jk}\delta_{mn}
\eeq
Since $\ga$ squares to one, it is invertible and the above mapping
is one-to-one. Let
\beq 
n_{jm}\equiv a_{jm}^\dagger a_{jm}
\eeq
be the number density operator for the species $j$.
One can then express the matrix $\ga$ at site $m$ in terms 
of fermions operators
\beq
\ga_m=\prod_{i=1}^d (2n_{im}-1)
\eeq
Since $n_{jm}$ squares to one $\ga_m$ also squares one as needed,
and the inverse equations are given by
\beq
\ga^+_{jm} = \prod_{k=1}^{m-1} \left(\prod_{l=1}^d (2 n_{lk}-1)
\right) a_{jm}^\dagger \;\;,\;\; \ga^-_{jm}=
\prod_{k=1}^{m-1} \left(\prod_{l=1}^d (2 n_{lk}-1) \right)a_{jm} \;\;,\;\;\;j=1,...,d \label{invfor}
\eeq

Some remarks are in order. One could have chosen to 
group the $\gamma_i$ matrices two by two in an arbitrary order, with
an eventual simple change in the definition of the $\gamma$ matrix. 
However there is no loss of generality in the foregoing choice, as
no specific choice of representation of the algebra ${\cal C}_{2d}$ 
has yet been made.

As we noted earlier, one 
should not be left with the impression that the Hilbert  
space at every site must have dimension $2^d$ to be able to  fermionize.
Indeed any  space can be embedded in the nearest 
higher dimensional space of dimension $2^d$. More precisely, the operators
at this site can be written as linear combinations of $E^{ij}$
matrices whose only non-vanishing elements is a one at row $i$ and column $j$,
with $i$, $j$ varying from one to the dimension of the  space. 
One first embeds the matrices $E^{ij}$ in the nearest $2^d$-dimensional
Hilbert space by adding the required rows and columns of zero,
and  then proceeds with the fermionization.  The transformed quantities
act in the larger Hilbert space but also act
in a stable way when restricted to the original
space, where one has identified which states of the 
larger Fock space span the smaller one. 
One obtains in this way  fermionized quantities for the chain. 
We shall illustrate this procedure on two  simple examples. 

The above transformations admit several variants.
The spaces at different sites of a chain need not be equal or of equal
dimension. Indeed, all that is required for  all that was said above
to still hold  is the one-to-one mapping
at one site and the anticommutation relations. And this can be ensured
by replacing the string $\ga_1\cdots \ga_{m-1}$
in equations (\ref{dm},\ref{dm1}) with
\beq
\pi_{m-1}^\gamma=\ga_1^{(1)}\cdots\ga_{m-1}^{(m-1)}\label{string1}
\eeq
where $\ga_i^{(i)}$ is the special gamma matrix acting
on the local space  of dimension $2^{d_i}$ at  site $i$.
The  dimension $2^{d_i}$ depends on the site. 
The $\gamma_j^{\pm}$ also correspond to the dimension $2^{d_i}$. 
Thus one can fermionize for instance a spin-chain with different 
representations at every site.  

One can also choose to fermionize selectively, leaving
some sites unfermionized while fermionizing others.
This is achieved through another change in $\pi_{m-1}^\gamma$.
For the sites where one keeps the bosonic variables, the 
bosonic operators are left unchanged by the mapping: 
$B_i \rightarrow B_i$. For the sites fermionized,
one uses another version of (\ref{string1}) where the string 
has been replaced by one where  the identity replaces
the $\gamma$-matrix  for all the sites
{\it not\/} fermionized. One thus has ensured that the fermionic operators
anticommute among themselves as needed, while they commute
with the bosonic operators for sites where no fermionization has taken place.
In this way one gets a chain of mixed bosonic and fermionic local spaces.

There is also the multiple-copies fermionization, when a local
space is a tensor product of smaller spaces.
The fermionization formulae are just the $\gamma$-generalizations
of  formulae (\ref{multiple}). One replaces the $p$ copies
of Pauli matrices by $p$ copies of gamma matrices. Note
that copies of  gamma matrices at the {\it same} site need not
correspond to the same dimension, as the space at a given site
may be the tensor product of spaces of different dimensions. 
Again, all the above variants  can still be implemented within 
the different
copies, with copies of gamma matrices at different sites 
possibly having different dimensions. The selective fermionization
is  still possible. 
We shall give examples of fermionizations with two copies of  
$\gamma$ matrices.  
Finally, reverse fermionization {\it i.e.\/} bosonization, with all its
variants, can be implemented  by use of the inversion formulae (\ref{invfor}). 
    
For the specific calculations to follow we have adopted 
the definitions of $\gamma$-matrices found in \cite{zinn}.
We now fermionize operators of various spin chains.

\section{Integrable spin-chains}

We start with a well-studied  model, the spin-$\frac{1}{2}$ XXZ
spin-chain. Its Hamiltonian as well as 
its integrable structure have  already been fermionized.
This simple example is a warm-up exercise which 
allows us to introduce the general framework of
integrability and general aspects of the fermionization
procedure.   We shall then fermionize 
the  simplest XXC models. 
Unless otherwise indicated, roman symbols  correspond 
to the original bosonic
quantities, while calligraphic ones correspond to the fermionized ones.

\subsection{The spin-$\frac{1}{2}$ XXZ chain}

We first consider the Hamiltonian and the $(L,R)$-system, and
then  boundary and symmetry issues.

\subsubsection{Hamiltonian and $L$-matrix}

The Hamiltonian is given by
\bea
H_2 &=&\sum_m \left(x^{-1}\,\sig^+_m \sig^-_{m+1} +x \sig^-_m \sig^+_{m+1}+
\Delta\,\sig^z_m \sig^z_{m+1}\right)\label{h11}\\
&=&\frac{1}{2}\sum_m \left(\pm\sig^x_m \sig^x_{m+1}\pm\sig^y_m 
\sig^y_{m+1}+ 2\Delta \,\sig^z_m \sig^z_{m+1}\right) \quad {\rm for} \;\; 
x=\pm 1\nonumber
\eea
Here we have introduced a twist parameter $x$. In the notation
of \cite{xxc} one is considering the model $(n_1=1,n_2=1)$.
The parameter   $x$   
arises  from threading the ring-chain by a flux $N\phi$.
One then have  $x=e^{i e\phi(\vec{x}_m)}$
where  
\beq
\phi(\vec{x}_m)=\int_{\vec{x}_m}^{\vec{x}_{m+1}} \vec{A}(\vec{x})d\vec{x}
\eeq
is the Peierls phase. 

The Jordan-Wigner transformation yields
\beq
\CH_2 =-\sum_m \left( x\,a^\dagger_{m+1} a_m +x^{-1} 
a^\dagger_m a_{m+1} -\Delta \,(2n_m-1)(2n_{m+1}-1) \right) \label{fh11}
\eeq
where $\Delta\equiv -\frac{\cos\gamma}{2}$, and 
$n_m\equiv a^\dagger_m a_m$ is the number density operator.  
Let us also introduce another number density operator,  
\beq
\nbar_m\equiv a_m a^\dagger_m=1-n_m\; , \quad n_m\,\nbar_m=\nbar_m\, n_m = 0
\eeq 
as this quantity
appears naturally farther on. 
Note that the transformation  $a_m\longleftrightarrow a^\dagger_m$
interchanges  $n_m$ and $\nbar_m$. This is the particle-hole
symmetry which reflects the fact that the roles of the  creation
and annihilation operators can be reversed. Since operators are
independent of the choice of a fermionic vacuum $a$ and $a^\dagger$
play symmetrical roles,  and it is only natural that $\nbar_m$ as well
as $n_m$ should appear.

The bosonic and fermionic  models, both
with their respective periodic boundary conditions,
are  known to be integrable. 
The fermionization of the $L$-$R$ 
system was done  in \cite{puzh}.
The fermionization of the integrable structure consists in applying 
a `gauge transformation' on the Lax matrix at every site.
The $R$-matrix intertwining two such matrices is then obtained 
by plugging the fermionic $L$-matrix into  the $RLL$ equation and
by cancelling out the transformation matrices.

We adopt the notations of \cite{xxc} and consider a 
transformation for $L$ which differs slightly 
from the one found in \cite{puzh,woa};
the difference is however not essential.
Let $a=\sin(\gamma - \lambda)$, 
$b=\sin\lambda$, $c=\sin\gamma$, and $E^{\af\be}$ be the matrix whose
sole non-vanishing entry is a one at row $\af$ and column $\be$. 
The $L=R$ matrix is given by:
\beq
L_m=       \left(\begin{array}{cc}
             a\,E^{11}_m+x\, b\,E^{22}_m & c\,E^{21}_m \\
            c\,E^{12}_m& x^{-1} b\,E^{11}_m +a\,E^{22}_m 
       \end{array} \right)
\end{equation}
Define the operators
\begin{equation}
V_m={\rm diag}(v_m,v_m^{-1})=  \left(\begin{array}{cc}
            v_m & 0  \\
            0 & v_m^{-1}
       \end{array} \right)
\quad , \quad v_m=\exp\left(\frac{i\pi}{2} \sum_{j=1}^{m-1}(n_j-1)\right)
 \end{equation}
The fermionic $\CL$-matrix is obtained as:
\beq
\CL_m = V_{m+1} L_m V_m^{-1} \label{gt1}
\eeq
and one easily derives:
\begin{equation}
\CL_m=       \left(\begin{array}{cc}
            a \, n_m-  i\,b\,x \,\nbar_m & -i\,c\,a_m  \\
            c\,{a^\dagger}_m &  i\,a\,\nbar_m+b\,x^{-1} n_m 
       \end{array} \right)\label{clxxz}
\end{equation}
The $\check{R}LL$ equation is given by
\beq
\check{R}(\la_1-\la_2) \; L(\la_1)\otimes L(\la_2) = L(\la_2)\otimes L(\la_1)
\;\check{R}(\la_1-\la_2)\label{rllc}
\eeq
and $\check{R}$ satisfies the Yang-Baxter equation
\beq
\check{R}_{12}(\la_1-\la_2)  \check{R}_{23}(\la_1)\check{R}_{12}(\la_2) =
\check{R}_{23}(\la_2) \check{R}_{12}(\la_1)
\check{R}_{23}(\la_1-\la_2)\label{ybec}
\eeq
Let $P$ be the permutation operator on the tensor product of
two  spaces: $P(x\otimes y) =y\otimes x$, where $x$ and $y$ are 
two vectors. Define $R(\la)\equiv P \check{R}(\la)$, then an equivalent form 
of equation (\ref{rllc}) is:
\beq
R(\la_1-\la_2) \; \stackrel{1}{L}(\la_1) \stackrel{2}{L}(\la_2) 
= \stackrel{2}{L}(\la_2) \stackrel{1}{L}(\la_1)
\;R(\la_1-\la_2)\label{rll}
\eeq
where $\stackrel{1}{L}(\la_1)=L(\la_1)\otimes\bI$ and
$\stackrel{2}{L}(\la_2)= \bI \otimes L(\la_2)$.
The Yang-Baxter equation for $R$ is given by
\beq
R_{12}(\la_1-\la_2)  R_{13}(\la_1) R_{23}(\la_2) =
R_{23}(\la_2) R_{13}(\la_1) R_{12}(\la_1-\la_2)\label{ybe}
\eeq

To find the fermionic version of (\ref{rllc}) one inverts (\ref{gt1})  
to obtain $L(\la_1)$ and $L(\la_2)$, and replaces
them into equation (\ref{rllc}). Upon commuting
the $V$'s  through, one finds that they  cancel 
each other out after leaving phases which modify the matrix $\check{R}$.
Equation  (\ref{rllc}) becomes
\beq
\check{\CR}(\la_1-\la_2) \; \CL(\la_1)\ots \CL(\la_2) = 
\CL(\la_2)\ots \CL(\la_1)\;\check{\CR}(\la_1-\la_2)\label{frllc}
\eeq
where the tensor product of two operators $A$ and $B$ 
is graded according to 
\beq
(A\ots B)_{ij,kl}= (-1)^{(P(i)+P(k))P(j)}
A_{ik} B_{jl} \quad , \quad P(i) \in \bZ_2
\eeq
and 
\beq
\check{\CR}(\la)=F^{-1}\check{R}(\la) F \;\;,\;\;\; 
F={\rm diag}(1,1,i,i)= \left({\rm diag}(1,i)\right)\otimes \bI_2
\eeq
For (\ref{clxxz}) one has: $P(1)=0$ and  $P(2)=1$.
Note that $F$ is a diagonal matrix  defined up to an overall normalization. 
We shall choose  throughout this paper $F_{11}=1$. This gives the above
$F$, and   $F^4=\bI_4$.
 
One can then define a graded permutation operator $\CP$ whose action
on two vectors of well-defined parity is given by:
$\CP (x\ots y)=(-1)^{P(x)P(y)}\,  y \ots x$.
In components one has: $\CP_{ij,kl} = (-1)^{P(i)P(j)}\delta_{il}\delta_{jk}$.
Let $\CR=\CP \check{\CR}$, and 
equation (\ref{rll}) becomes
\beq
\CR(\la_1-\la_2) \; \stackrel{1}{\CL}(\la_1) \stackrel{2}{\CL}(\la_2) 
= \stackrel{2}{\CL}(\la_2) \stackrel{1}{\CL}(\la_1)
\;\CR(\la_1-\la_2)\label{frll}
\eeq
where $\stackrel{1}{\CL}(\la_1)=\CL(\la_1)\ots\;\bI$ and
$\stackrel{2}{\CL}(\la_2)= \bI\; \ots \CL(\la_2)$.
The matrices $\check{\CR}$ and  $\CR$ satisfy the Yang-Baxter equations 
(\ref{ybec}) and  (\ref{ybe}) respectively.
However, while (\ref{ybec}) written in components for $\check{\CR}$ 
is unchanged, equation (\ref{ybe}) for $\CR$
has grading signs due to the  graded permutation operator.

We shall obtain similar results for the fermionization
of all the integrable systems considered here. The above equations 
will still hold, with the corresponding matrices $R$, $L$, 
$F$,  and the appropriate grading. 

\subsubsection{Boundary terms and integrability}

It is well-known that the Jordan-Wigner transformation  does 
not conserve periodic boundary conditions. One finds
\beq
\sig_N^+ \sig_1^- =-(2n_2-1)\cdots(2n_{N-1}-1) a^\dagger_N a_1
\;\;,\;\; \sig_N^- \sig_1^+ =-(2n_2-1)\cdots(2n_{N-1}-1) a^\dagger_1 a_N
\eeq
Thus the fermionic Hamiltonian (\ref{fh11}) with the JW-twisted fermionic
boundary conditions
\beq
a_{N+1}\equiv  +(2n_2-1)\cdots(2n_{N-1}-1)a_1 \;\;,\;\;
a^\dagger_{N+1}\equiv  +(2n_2-1)\cdots(2n_{N-1}-1) a_1^\dagger
\label{jwtwist}
\eeq
is the Hamiltonian equivalent to the bosonic one (\ref{h11})
with periodic boundary conditions.   
If instead one takes  periodic fermionic boundary conditions
for (\ref{fh11}), with $a_{N+1}\equiv a_1$ and $a^\dagger_{N+1}\equiv
a^\dagger_1$, the resulting Hamiltonian is not anymore equivalent
to the bosonic one with periodic boundary conditions. It is however
{\it still\/} integrable. This follows from the fermionic 
relations (\ref{frllc}). 

To this end  we recall
the general tenets of integrability. 
The transfer matrix, $\tau(\la)$, 
is the generating functional of the infinite set 
of conserved quantities.
Its construction in the framework of the Quantum Inverse Scattering 
Method (QISM) is well known \cite{qism1,qism2,kbi}.
Given an $(L,R)$ pair,  the trace over 
the auxiliary space of the monodromy matrix $T(\la)$ yields the
transfer matrix:  
\beq
\tau (\la)\equiv {\rm Tr}_0 \;\left[T(\la)\right]\equiv
{\rm Tr}_0 \;\left[M_0\, L_{0N}(\la)\cdots L_{01}(\la)\right]\label{transfy}
\eeq
where $N$ is the number of sites on the chain and 0 is the auxiliary
space.
The introduction of the numerical matrix 
$M$ corresponds to integrable periodic $M$-twisted 
boundary conditions, with $M=\bI$ corresponding to periodic conditions. 
This twisting of the periodic boundary 
conditions is different  from the one arising from the Jordan-Wigner 
transformation. 
For matrices $M$ such that ${[M\otimes M, \check{R}(\la)]}=0$
equation (\ref{rllc}) implies the $RTT$ relations: 
\beq
\check{R}(\la_1-\la_2) \; T(\la_1)\otimes T(\la_2) =  T(\la_2)\otimes T(\la_1)
\;\check{R}(\la_1-\la_2)\label{rtt}
\eeq
Taking the trace over the auxiliary spaces 
one obtains $[\tau(\la_1),\tau(\la_2) ]=0$.
A set of local conserved quantities is  given by
\beq
H_{p+1} = \left({d^p \ln\tau (\la)\over d\lambda^p}\right)_{\la=0}
\;\;\; , \;\; p\geq 0 \label{cqs}
\eeq 
The Hamiltonians $H_{p+1}$ mutually commute and the system is said 
to be integrable. 

In the fermionic setting 
the monodromy matrix is given by
\beq
\CT(\la)\equiv M_0\, \CL_{0N}(\la)\cdots\CL_{01}(\la)
\eeq
This is so because
because the Lax matrix is a homogeneous even-parity matrix.
The generalization of the QISM framework to the graded context
was initiated in \cite{ks, ppk}.  
Equation (\ref{frllc}) replaces 
equation (\ref{rllc}), with the appropriate grading, and 
equation (\ref{rtt}) is replaced by
 \beq
\check{\CR}(\la_1-\la_2) \; \CT(\la_1)\ots \CT(\la_2) 
=  \CT(\la_2)\ots \CT(\la_1) \;\check{\CR}(\la_1-\la_2)\label{frtt}
\eeq
It can be  shown in all generality 
that periodic fermionic boundary conditions correspond 
to a matrix $M$ which  implements a supertrace:
\beq
M_{ij}= \delta_{ij} (-1)^{P(i)}\quad ,\quad
{\rm Tr} (M A)  ={\rm Str} (A) = \sum_i (-1)^{P(i)} A_{ii}
\eeq
This shows that Hamiltonian (\ref{fh11}) with fermionic periodic 
boundary conditions is also integrable.

We restrict ourselves to $M=\bI$ for the bosonic case,
and to a matrix implementing the supertrace in the fermionic setting.
The latter matrix corresponds to the grading arising in equation (\ref{frllc}).
For the latter case we omit $M$ and replace the trace by a supertrace
in (\ref{transfy}). 
From now on, unless otherwise indicated,  `bosonic system' means we consider
bosonic periodic boundary conditions, while  `fermionic system'
means we consider  periodic fermionic boundary conditions.
Thus we are considering the bosonic and fermionic systems
with their `natural' boundary conditions; as we have seen,
these systems are {\it inequivalent}.

An immediate and general  consequence of all this 
is the existence of {\it at least\/} two integrable boundary conditions for
any integrable system: Periodic and possibly many periodic-twisted
conditions  through
fermionization or bosonization. This applies to the general 
fermionization scheme we introduced.  

\subsubsection{Symmetries}

Let us now consider the symmetries of the 
bosonic spin-$\frac{1}{2}$ XXZ model. 
For arbitrary $\Delta$ the operator $\sig^z=\sum_{i=1}^N \sig^z_i$
commutes with the transfer matrix  and therefore with all the conserved 
quantities. At the rational point where
$\Delta=\frac{1}{2}$, the full symmetry is $su(2)$,
with the two remaining generators being $\sig^+=\sum_{i=1}^N \sig^+_i$
and $\sig^-=\sum_{i=1}^N \sig^-_i$. 
Their fermionic counterparts are given by:
\beq
\sig^+ = \sum_{i=1}^N \left(\prod_{j=1}^{i-1} (2n_j-1)\right)
 a^\dagger_i \; ,\;\;
\sig^- = \sum_{i=1}^N \left(\prod_{j=1}^{i-1} (2n_j-1)\right) a_i\; ,\;\;
\sig^z = \sum_{i=1}^N  (2n_i-1)
\eeq
To show that $\sig^z$ commutes with the fermionic 
transfer matrix, we showed and used
\beq
{[n_m,\CL_m(\la)]}=-\frac{1}{2} {[\sig_0^z, \CL_m(\la)]}
\eeq
At $\Delta=\frac{1}{2}$, are  $\sig^{\pm}$ still  symmetries
of the periodic fermionic system? We have verified that
the fermionic Hamiltonian does {\it not\/} commute
with  $\sig^{\pm}$. Thus, although 
the rational limit of $\CL$ and $\check{\CR}$   is well-defined, 
the symmetry enhancement at the rational point
does not take place for the fermionic system. 
This is in fact a general (negative) 
result for all models fermionized, whenever  the
rational limit provides an enlarged symmetry for the bosonic system. 

This simple fermionization
already illustrate features that will also arise for 
most if not all integrable systems.

\subsection{The $(1,2)$-XXC model}\label{3dim}

This models naturally appears when one takes the infinite coupling limit
of the Hubbard model; see for instance \cite{aars}. 
The dimension of the the Hilbert space at one site is three. 
We therefore utilize the $\CC_4$  Clifford algebra to fermionize this 
model. 

The Hamiltonian  is given by
\beq
H_2= \sum_m \left(\sum_{\be=2}^3( x_{1\be} E^{\be 1}_m E^{1\be}_{m+1} 
+ x_{1\be}^{-1} E^{1\be}_m E^{\be 1}_{m+1}) - \cos\ga \,
(E^{11}_m E^{11}_{m+1} + \sum_{\be,\be^{'} =2}^3
E^{\be\be}_m E^{\be^{'}\be^{'}}_{m+1})\right)
\eeq
We took $A=\{ 1\}$ and $B=\{ 2,3\}$ in the notation of \cite{xxc}. 
In the bosonic
setting the two other choices 
are unitarily equivalent to this one \cite{mm}.

There are four possible equivalent embeddings of the $3\times 3$ matrices
into $4\times 4$ matrices:
\beq
\begin{array}{ll}
(1,2,3) \rightarrow (1,2,3,4)\;,\;\; &
(1,2,3) \rightarrow (1,2,4,3)\;,\;\;\\
(1,2,3) \rightarrow (1,3,4,2)\;,\;\; &
(1,2,3) \rightarrow (2,3,4,1)\;.
\end{array}
\eeq
For instance, the second choice corresponds to 
letting
\beq
\left(\begin{array}{ccc}
            a_{11} &  a_{12} & a_{13}  \\
            a_{21} &  a_{22} & a_{23}  \\
            a_{31} &  a_{32} & a_{33}
       \end{array} \right)
\longrightarrow 
\left(\begin{array}{cccc}
            a_{11} &  a_{12} & 0 & a_{13}  \\
            a_{21} &  a_{22} & 0 & a_{23}  \\
            0      &  0      & 0 & 0       \\
            a_{31} &  a_{32} & 0 & a_{33}
       \end{array} \right)
\eeq
while the fourth choice corresponds to
\beq
\left(\begin{array}{ccc}
            a_{11} &  a_{12} & a_{13}  \\
            a_{21} &  a_{22} & a_{23}  \\
            a_{31} &  a_{32} & a_{33}
       \end{array} \right)
\longrightarrow 
\left(\begin{array}{cccc}
            0 &   0    &    0    &   0     \\
            0 & a_{11} &  a_{12} & a_{13}  \\
            0 & a_{21} &  a_{22} & a_{23}  \\
            0 & a_{31} &  a_{32} & a_{33}
       \end{array} \right)
\eeq
In the  specific representation we adopted for the $\gamma$-matrices,
the canonical basis $(e_1,e_2,e_3,e_4)$ is sent to the
four fermionic states as follows:
\beq
\begin{array}{ll}
e_1=\left( \begin{array}{c}
1\\0\\0\\0 \end{array}\right)\longrightarrow a_1^\dagger a_2^\dagger |0\rangle
\quad&\quad
e_2=\left( \begin{array}{c}
0\\1\\0\\0\end{array}\right)\longrightarrow  |0\rangle \\
\\
e_3=\left( \begin{array}{c}
0\\0\\1\\0\end{array}\right)\longrightarrow a_1^\dagger  |0\rangle\quad &\quad
e_4=\left( \begin{array}{c}
0\\0\\0\\1\end{array}\right)\longrightarrow  a_2^\dagger |0\rangle
\end{array}
\eeq
We choose, for obvious reasons 
but  again without  loss of generality,
the fourth embedding which means that the Hamiltonian and the
higher conserved quantities, $\CH_p$, automatically annihilate all 
states of the chain which  have a double occupancy on at least one site.
The third choice corresponds to the annihilation
of all states which have a vacuum on  at least one site. 

The fermionized Hamiltonian density is given by
\bea
\CH_{mm+1} &=& (x_{23}\, a_{1m}^\dagger a_{1m+1}
+x_{23}^{-1} a_{1m+1}^\dagger a_{1m}) \nbar_{2m}\nbar_{2m+1}\\
&+& (x_{24}\, a_{2m}^\dagger a_{2m+1}
+x_{24}^{-1} a_{2m+1}^\dagger a_{2m}) \nbar_{1m}\nbar_{1m+1}\nonumber\\
&-&\frac{1}{2} \cos\gamma\; ( 1+\CC_m \CC_{m+1})\nonumber
\eea
where $\CC= (2n_1-1)(2n_2-1)-n_1 n_2$ is the `conjugation operator' \cite{ff}.
It is obvious that the action of  $\CH_2$ is stable on the subspace 
of dimension $3^N$.
Define the projection operator which annihilates
any state with double occupancy on at  least one site:
\beq
p=\prod_{m=1}^N (1-n_{1m} n_{2m})\label{proj}
\eeq
$\CH_2$ can be interpreted as acting on the reduced Hilbert space
or as $p \CH_2 p$ acting on the full $4^N$-dimensional space. 

The Lax matrix is fermionized 
with the help of the matrix
\beq
V_m= {\rm diag}(v_m,v_m^{-1},v_m^{-1})  \quad {\rm where}\quad 
v_m= \exp\left( \frac{i\pi}{2}\sum_{i=1}^{m-1} \sum_{j=1}^2 (n_{ji}-1)\right)
\eeq
One easily finds
\bea
\CL_m &=& \left(\begin{array}{ccc}
\CL_{22}&-i\,c\, a_{1m}^\dagger\nbar_{2m}&i\,c\, a_{2m}^\dagger\nbar_{1m}\\
-c\, a_{1m}\nbar_{2m}&\CL_{33}&-i\,a\, a_{2m}^\dagger a_{1m}\\
c\,a_{2m}\nbar_{1m}&-i\, a\, a_{1m}^\dagger a_{2m}& \CL_{44}
\end{array}\right)\\
& & \nonumber\\
\CL_{22} &=& -a\, \nbar_{1m}\nbar_{2m} -i\,b( x_{23}\, n_{1m}\nbar_{2m}
+x_{24}\, \nbar_{1m} n_{2m})  \nonumber\\
\CL_{33} &=&   i\,a\, n_{1m}\nbar_{2m} -\,b\, x_{23}^{-1} \nbar_{1m}\nbar_{2m}
\nonumber\\
\CL_{44} &=&   i\,a\, \nbar_{1m}n_{2m} -\,b\, x_{24}^{-1} \nbar_{1m}\nbar_{2m}
\nonumber
\eea
Note that $(1-n_{1m} n_{2m})\CL_m = \CL_m (1-n_{1m} n_{2m}) =\CL_m$. 
The grading is given by: $P(2)=0$, $P(3)=P(4)=1$. 
We obtain for the matrix $F$ 
\beq
F={\rm diag}(1,1,1,-i,-i,-i,-i,-i,-i)={\rm diag}(1,-i,-i)\otimes \bI_3
\eeq

The symmetry generators for the conserved quantities
are given by
\beq
\begin{array}{lccl}
X^{34}=\sum_m a_{2m}^\dagger a_{1m}^\dagger &  &  & X^{43}=\sum_m
a_{1m} a_{2m}\\
X^1= \sum_m n_{1m} &  &  & X^2= \sum_m n_{2m}
\end{array}
\eeq
Recall that the off-diagonal generators are symmetries provided
the following constraint hold: $x_{23}=x_{24}$ \cite{xxc}.
We see in the next section how to show similar commutation relations 
with the fermionic transfer matrix. 
The symmetries are the generators of $su(2)\times u(1)$, and
reflect a (partial) symmetry between the two species of spinless fermions.
Because of no-double-occupancy, $X=\sum_m n_{1m} n_{2m}$ is a trivial
symmetry.
Note that, here and below,
these fermions can be seen as one type of fermion with 
spin $\frac{1}{2}$. We however reserve the terms `up' and `down' for
the two XX copies used in constructing Hubbard models.

\subsection{The $(2,2)$-XXC model: first avatar}

We first write this XXC model with $A=\{1,2\}$ and $B=\{3,4\}$. 
The fermionic  Hamiltonian density is given  by 
\begin{eqnarray}
\CH_{mm+1}^{(1)}&=& a^\dagger_{1m}a_{1m+1}\left[x_{23}\,\nbar_{2m}
\nbar_{2m+1}-x^{-1}_{14}n_{2m}n_{2m+1}\right]\label{fh22}\\ 
          &+&a^\dagger_{1m+1}a_{1m}\left[ x^{-1}_{23}\nbar_{2m}
\nbar_{2m+1}-x_{14}\, n_{2m}n_{2m+1}\right] \nonumber\\ 
          &+&a^\dagger_{2m}a_{2m+1}\left[ x_{24}\,\nbar_{1m}
\nbar_{1m+1}-x^{-1}_{13}n_{1m}n_{1m+1}\right]\nonumber \\ 
          &+&a^\dagger_{2m+1}a_{2m}\left[ x^{-1}_{24}\nbar_{1m}
\nbar_{1m+1}-x_{13}\, n_{1m}n_{1m+1}\right] \nonumber\\ 
          &-&\frac{1}{2}\cos\gamma\; (1+\CC_m^{(1)} \CC_{m+1}^{(1)})\nonumber
\end{eqnarray} 
where $\CC^{(1)}=(2n_1-1)(2n_2-1)$.

The Lax matrix is fermionized 
with the help of matrix
\beq
V_m= {\rm diag}(v_m,v_m,v_m^{-1},v_m^{-1})\quad {\rm where}\quad 
v_m= \exp\left(\frac{i\pi}{2}\sum_{i=1}^{m-1}\sum_{j=1}^2 (n_{ji}-1)\right)
\eeq
We find 
\begin{equation}
\CL_m^{(1)}=   \left(\begin{array}{cccc}
    \CL_{11}^{(1)}&-a\,a_{1m}a_{2m}&-i\,c\,a_{2m}n_{1m}
&-i\,c\,a_{1m}n_{2m}\\
    -a\,a^\dagger_{1m}a^\dagger_{2m} & \CL_{22}^{(1)}&-i\,c\, 
a^\dagger_{1m}\bar{n}_{2m}&i\,c\,a^\dagger_{2m}\bar{n}_{1m}\\
            c\,a^\dagger_{2m}n_{1m}&-c\,a_{1m}\bar{n}_{2m}
&\CL_{33}^{(1)}&-i\,a\, a^\dagger_{2m}a_{1m}\\
            c\,a^\dagger_{1m}n_{2m}&c\,a_{2m}\bar{n}_{1m}
&-i\,a\,a^\dagger_{1m}a_{2m}&\CL_{44}^{(1)}
\end{array} \right)\label{l22xxc}
\end{equation}
\beas
&\CL_{11}^{(1)}& = a\,n_{1m}n_{2m}-i\,b\left\{ x_{13}\,n_{1m}
\nbar_{2m}+x_{14}\,\nbar_{1m}n_{2m} \right\} \\ 
&\CL_{22}^{(1)}& = -a\,\nbar_{1m}\nbar_{2m}-i\,b
\left\{ x_{23}\,n_{1m}\nbar_{2m}+x_{24}\,\nbar_{1m}n_{2m} \right\}\\ 
&\CL_{33}^{(1)}& = i\,a\,n_{1m}\nbar_{2m}+b\left\{ x^{-1}_{13}n_{1m}
n_{2m}-x^{-1}_{23}\nbar_{1m}\nbar_{2m} \right\} \\ 
&\CL_{44}^{(1)}& = i\,a\,\nbar_{1m}n_{2m}+b\left\{ x^{-1}_{14}
n_{1m}n_{2m}-x^{-1}_{24}\nbar_{1m}\nbar_{2m} \right\} 
\eeas
The tensor product of equation (\ref{frllc})  is now graded according to
\beq
P(1)=P(2)=0\;,\;\; P(3)=P(4)=1
\eeq
For the matrix $F$ we found
\beq
F={\rm diag}(1,1,1,1,-1,-1,-1,-1,i,i,i,i,i,i,i,i)
= {\rm diag}(1,-1,i,i)\otimes \bI_4
\eeq
Again one has $F^4=\bI_{16}$. 

Now consider  the symmetries of the transfer matrix.
It is easy to check that the following commutation relations hold
provided all the parameter $x_{a\be}$ are equal to each 
other, but not necessarily to $\pm 1$:
\beq
\begin{array}{lccccl}
\{ a_{2m}^\dagger a_{1m}^\dagger, \CL_m^{(1)}\} ={[E_0^{12},\CL_m^{(1)}]}& 
& & & &
\{ a_{1m} a_{2m}, \CL_m^{(1)}\} ={[E_0^{21},\CL_m^{(1)}]}\\
{[ a_{2m} a_{1m}^\dagger, \CL_m^{(1)} ]} =-{[E_0^{34},\CL_m^{(1)}]}& 
& & & &
{[ a_{1m} a_{2m}^\dagger, \CL_m^{(1)} ]} =-{[E_0^{43},\CL_m^{(1)}]}\\
{[ n_{1m}n_{2m}, \CL_m^{(1)} ]} =-{[E_0^{11},\CL_m^{(1)}]}& & & & &
{[ \nbar_{1m}\nbar_{2m}, \CL_m^{(1)} ]} =-{[E_0^{22},\CL_m^{(1)}]}\\
{[ n_{1m} \nbar_{2m}, \CL_m^{(1)} ]} =-{[E_0^{33},\CL_m^{(1)}]}& & & & &
{[ \nbar_{1m} n_{2m}, \CL_m^{(1)} ]} =-{[E_0^{44},\CL_m^{(1)}]}
\end{array}\label{comr}
\eeq
Let
\beq
\begin{array}{lccccl}
X^{12}=\sum_m a_{2m}^\dagger a_{1m}^\dagger &  &  & & & 
X^{21}=\sum_m a_{1m} a_{2m} \\
X^{34}=\sum_m a_{2m} a_{1m}^\dagger &  &  & & &
X^{43}=\sum_m a_{1m} a_{2m}^\dagger\\
X^{11}=\sum_m n_{1m} n_{2m}&  &  & & &
X^{22}=\sum_m \nbar_{1m} \nbar_{2m}\\
X^{33}=\sum_m n_{1m} \nbar_{2m}&  &  & & &
X^{44}=\sum_m \nbar_{1m} n_{2m}
\end{array}\label{syme}
\eeq
Relations (\ref{comr}) imply the following commutation relations
with the  transfer matrix of the fermionic system:
\beq
\begin{array}{lccccl}
\{ X^{12}, \tau^{(1)}(\la)\} = 0&  &  & & &
\{ X^{21}, \tau^{(1)}(\la)\} = 0\\
{[ X^{34}, \tau^{(1)}(\la)]} = 0&  &  & & &
{[ X^{43}, \tau^{(1)}(\la)]} = 0\\
{[ X^{11}, \tau^{(1)}(\la)]} = 0&  &  & & &
{[ X^{22}, \tau^{(1)}(\la)]} = 0\\
{[ X^{33}, \tau^{(1)}(\la)]} = 0&  &  & & &
{[ X^{44}, \tau^{(1)}(\la)]} = 0
\end{array}
\eeq
We come to the unusual fact that the zero parity operators
$X^{12}$ and $X^{21}$ {\it anticommute\/} with the zero parity operator
$\tau(\la)$ ! However,
a simple proof by induction shows that the logarithmic 
derivatives of the transfer matrix are made up of sums
of product of an even number of the transfer matrix and its derivatives.
This shows that the symmetries $X^{12}$ and $X^{21}$ 
still commute with the conserved
quantities $\CH_p$, for $p\geq 1$.
The anticommutation  with $\exp(\CH_1)$ does not of course prevent
the   simultaneous diagonalization of $X^{12}$ or $X^{21}$ 
with all the other 
conserved quantities,
as the spectral spaces are stable under anticommutation 
and/or commutation relations. 
The symmetry operators clearly are the generators of 
$su(2)\times su(2) \times u(1)$, with a fermionic realization
of the generators of this algebra. 

\subsection{The $(2,2)$-XXC model: second avatar}\label{avatar2}

We  consider here the $(2,2)$-XXC model in another realization,
where $A=\{ 1,3 \}$ and $B=\{ 2,4 \}$.
As first pointed out  in \cite{mm}, this bosonic model is unitarily equivalent
to the bosonic model of the preceding section.
The orthogonal operator linking 
the second avatar to the first one is constructed as follows.
Consider the orthogonal matrix
\beq
U=U^{-1}=U^t=\left(\begin{array}{cccc}
1&0&0&0\\0&0&1&0\\0&1&0&0\\0&0&0&1
\end{array} \right)
\eeq
The  $N$-fold tensor product $U^{(N)}=U\otimes \cdots \otimes U$
transforms the models into each other, modulo redefinitions of the 
parameters $x_{a\be}$. This can be seen at the Hamiltonian level 
or generally at the transfer matrix level
\beq
U^{(N)}\, \tau^{(1)}(\la)\, U^{(N)} = \tau^{(2)}(\la) \quad\Longrightarrow
\quad U^{(N)}\, H_p^{(1)} U^{(N)} = H_p^{(2)}  
\eeq
At the level of the  $L$-matrices one has:
\beq
L^{(2)}_m= U_0\,U_m\, L^{(1)}_m U_m U_0
\eeq
Note that $U^{(N)}$ is an orthogonal and therefore unitary operator. 

The fermionized Hamiltonian density  is now given by:
\begin{eqnarray}
\CH_{mm+1}^{(2)}&=& a^\dagger_{1m}a_{1m+1}\left[ 
x^{-1}_{32}\nbar_{2m}\nbar_{2m+1}-x^{-1}_{14}n_{2m}n_{2m+1}\right]\\
&+& a^\dagger_{1m}a_{1m+1}\left[x^{-1}_{12}a^\dagger_{2m}a_{2m+1}-x^{-1}_{34}
a^\dagger_{2m+1}a_{2m}\right] \nonumber\\ 
&+&a^\dagger_{1m+1}a_{1m}\left[ x_{32}\,\nbar_{2m}\nbar_{2m+1}
-x_{14}\,n_{2m}n_{2m+1}\right]\nonumber\\
&+&a^\dagger_{1m+1}a_{1m}\left[x_{12}\,a^\dagger_{2m+1}a_{2m}
-x_{34}\,a^\dagger_{2m}a_{2m+1}
\right] \nonumber\\
&-&\frac{1}{2} \cos\gamma\,(1+\CC_m^{(2)} \CC_{m+1}^{(2)})\nonumber
\end{eqnarray}
where $\CC^{(2)}=2 n_1-1$.
The $V$-matrix,  the induced grading and matrix $F$ are the same 
as in the previous section. 
We find:
\begin{equation}
\CL_m^{(2)}=   \left(\begin{array}{cccc}
            \CL_{11}^{(2)}&-c\, a_{1m}a_{2m}&-i\,a\, a_{2m}n_{1m}
&-i\,c\, a_{1m}n_{2m}\\
            -c\, a^\dagger_{1m}a^\dagger_{2m} & \CL_{22}^{(2)}&-i\,c\,
a^\dagger_{1m}\bar{n}_{2m}&i\,a\,a^\dagger_{2m}\bar{n}_{1m}\\
            a\,a^\dagger_{2m}n_{1m}&-c\,a_{1m}\bar{n}_{2m}&\CL_{33}^{(2)}
&-i\,c\, a^\dagger_{2m}a_{1m}\\
            c\,a^\dagger_{1m}n_{2m}&a\,a_{2m}\bar{n}_{1m}
&-i\,c\,a^\dagger_{1m}a_{2m}&\CL_{44}^{(2)}
       \end{array} \right)
\end{equation}
\beas
&\CL_{11}^{(2)}& = a\,n_{1m}n_{2m}-b\left\{ 
x_{12}\,\nbar_{1m}\nbar_{2m}+i\,x_{14}\,\nbar_{1m}n_{2m} \right\} \\ 
&\CL_{22}^{(2)}& = -a\,\nbar_{1m}\nbar_{2m}+b\left\{ 
x^{-1}_{12}n_{1m}n_{2m}-i\,x^{-1}_{32}n_{1m}\nbar_{2m} \right\} \\ 
&\CL_{33}^{(2)}& = i\,a\,n_{1m}\nbar_{2m}-b\left\{ 
x_{32}\,\nbar_{1m}\nbar_{2m}-i\,x_{34}\,\nbar_{1m}n_{2m} \right\} \\ 
&\CL_{44}^{(2)}& = i\,a\,\nbar_{1m}n_{2m}+b\left\{ 
x^{-1}_{14}n_{1m}n_{2m}+i\,x^{-1}_{34}n_{1m}\nbar_{2m} \right\} 
\eeas
This matrix is not fundamentally different from the one in the 
preceding section. 

Unlike the fermionic version of the first realization,
the symmetries which survive fermionization are only the diagonal
ones: the $X^{ii}$, $i=1,...,4$. They generate $u(1)\times u(1)\times u(1)$.
There are no constraints on the parameters $x_{a\be}$.
The off-diagonal bosonic symmetry generators become non-local in the 
fermionic setting  and it is easy 
to check that they do not commute with the fermionic Hamiltonian.  
This breaks the bosonic $su(2)\oplus su(2) \oplus u(1)$ to 
the three $u(1)$'s. 

The obstruction to symmetry comes from  
the periodic fermionic boundary conditions.
Locally the two fermionic versions are however equivalent. At the 
level of the $\CL$ matrices one has:
\beq
\CL^{(2)}_m= (V_{m+1} U_0\, \CU_m V_{m+1}^{-1}) \,\CL^{(1)}_m \,
(V_m U_0\, \CU_m V_m^{-1}) 
\eeq
The  operator $\CU_m$ is non-local:
\beq
\CU_m=n_{2m}+\left(\prod_{j=1}^{m-1} (2n_{1j}-1)(2n_{2j}-1)\right) \nbar_{2m} 
(a_{1m}^\dagger +a_{1m})
\eeq
At the Hamiltonian level the correspondence is:
\beq
\CH^{(2)}_{mm+1} =\CU_N \cdots \CU_1 \CH_{mm+1}^{(1)} \CU_1 \cdots \CU_N \;,
\quad\quad m=1,...,N-1
\eeq
However this equivalence breaks down for the last term: $\CH_{N1}$.
Thus the two fermionic models, despite sharing a common bosonic antecedent, are 
{\it not} equivalent. This is due to the periodic fermionic boundary conditions.

This is not the end of the story however. By fine tuning the twist
parameters one can recover a large superalgebra symmetry. 
Integrability is preserved since it does not depend on the values of these
parameters. We take
\beq
x\equiv x_{12}= x_{32}=-x_{34}=x_{14}
\eeq
The following operators then are symmetries of 
all the conserved quantities:
\beq
\begin{array}{lccccl}
Y^{11}=\sum_m n_{1m} n_{2m} & & & & & Y^{22}=\sum_m \nbar_{1m} \nbar_{2m}\\
Y^{33}=\sum_m n_{1m} \nbar_{2m} & & & & & Y^{44}=\sum_m \nbar_{1m} n_{2m}\\
Y^{13}=\sum_m n_{1m} a_{2m}^\dagger & & & & &  Y^{31}=\sum_m n_{1m} a_{2m}\\
Y^{24}=-\sum_m \nbar_{1m} a_{2m} & & & & &   Y^{42}=-\sum_m 
\nbar_{1m} a_{2m}^\dagger\label{syme1}
\end{array}
\eeq
They generate a fermionic realization of two
copies of a Lie superalgebra: $gl(1|1)\times gl(1|1)$.
The off-diagonal operators $Y$ appear as the `localized' counterparts
of the bosonic operators. 

\vskip 0.8cm
{\noindent
{\it The}} $(2,2)-XXC$ {\it model: third avatar}\hfill\\
There is also a $3^{\rm rd}$ avatar for this model
where one takes  $A=\{1,4\}$ and $B=\{2,3\}$. 
We found the quantities $\CH^{(3)}$, $\CL^{(3)}$ and verified that  
$V$, $F$ and the grading are unchanged. 
The symmetry breaks down to 
$u(1)\times u(1)\times u(1)$. However, upon localization of the fermionic
generators and fine tuning of the twist parameters, the symmetry is enlarged
to $gl(1|1)\times gl(1|1)$. Again, due to the boundary terms, 
this fermionic system is not equivalent to 
the previous two versions.
Going back to bosonic variables, we find three {\it integrable}
periodic-twisted  boundary conditions, in addition to the
starting bosonic periodic one.   
We say more about this in section (\ref{genrem}).

\subsection{The $(1,3)$-XXC model}

We now consider the second and last  XXC model with four states per site.
This model is  not equivalent to the above two models.
It has $su(3)\oplus u(1)$ symmetry. 
As we have seen on   the preceding two models,
a particular  choice of states for $A$ and $B$ is not restrictive,
at the bosonic level.
 
We take here $A=\{ 1,2,3 \}$ and $B=\{4 \}$. The fermionized Hamiltonian
density then reads:
\begin{eqnarray}
\CH_{mm+1}&=& -x^{-1}_{14}a^\dagger_{1m}a_{1m+1}n_{2m}n_{2m+1}
             -x_{14}\,a^\dagger_{1m+1}a_{1m}n_{2m}n_{2m+1} \\ 
           &+&x_{24}\,a^\dagger_{2m}a_{2m+1}\nbar_{1m}\nbar_{1m+1}
             +x^{-1}_{24}a^\dagger_{2m+1}a_{2m}\nbar_{1m}\nbar_{1m+1}  \nonumber\\
           &-&x^{-1}_{34}a^\dagger_{1m}a_{1m+1}a^\dagger_{2m+1}a_{2m}
             -x_{34}\,a^\dagger_{1m+1}a_{1m}a^\dagger_{2m}a_{2m+1} \nonumber\\
&-&\frac{1}{2} \cos\gamma\,(1+\CC_m \CC_{m+1})\nonumber
\end{eqnarray}
where $\CC=2 n_1 n_2 - 2 n_2+1$. 
The matrices $V$ and $F$, and the grading are again unchanged.  
The $L$-matrix reads:
\begin{equation}
\CL_m=   \left(\begin{array}{cccc}
            \CL_{11}&-a\, a_{1m}a_{2m}&-i\,a\,a_{2m}n_{1m}&-i\,c\, a_{1m}n_{2m}\\
            -a\,a^\dagger_{1m}a^\dagger_{2m} & \CL_{22}&-i\,a\,
a^\dagger_{1m}\bar{n}_{2m}&i\,c\,a^\dagger_{2m}\bar{n}_{1m}\\
            a\,a^\dagger_{2m}n_{1m}&-a\,a_{1m}
\bar{n}_{2m}&\CL_{33}&-i\,c\, a^\dagger_{2m}a_{1m}\\
            c\,a^\dagger_{1m}n_{2m}&c\,a_{2m}\bar{n}_{1m}&-i\,c\,
a^\dagger_{1m}a_{2m}&\CL_{44}
       \end{array} \right)
\end{equation}
\beas
&\CL_{11}& = a\,n_{1m}n_{2m}-i\,b\,x_{14}\,\nbar_{1m}n_{2m} \\ 
&\CL_{22}& = -a\,\nbar_{1m}\nbar_{2m}-i\,b\,x_{24}\,\nbar_{1m}n_{2m} \\
&\CL_{33}& = i\,a\,n_{1m}\nbar_{2m}+i\,b\,x_{34}\,\nbar_{1m}n_{2m} \\ 
&\CL_{44}& = i\,a\,\nbar_{1m}n_{2m}+b\left\{ x^{-1}_{14}n_{1m}n_{2m}
- x^{-1}_{24}\nbar_{1m}\nbar_{2m} +i\,x^{-1}_{34}n_{1m}\nbar_{2m} \right\} 
\eeas

For $x_{14}=x_{24}$ and the other $x$'s unconstrained,
the surviving symmetry is $su(2)\times u(1)\times u(1)$.
The generators are $X^{12}$ and $X^{21}$ and the four diagonal
ones in (\ref{syme}). If in additions one takes 
$x_{14}=x_{24}=-x_{34}$, one gets a fermionic realization of the
Lie superalgebra $gl(2|1)$, and an 
enlarged symmetry: $gl(2|1)\times u(1)$. The complete set of
generators is: $X^{12}$, $X^{21}$, $X^{11}$, $X^{22}$, $X^{33}$, $X^{44}$
of (\ref{syme}), $Y^{13}$, $Y^{31}$ of (\ref{syme1}), and
$Y^{23}=\sum_m \nbar_{2m} a_{1m}$, $Y^{32}=\sum_m \nbar_{2m} a_{1m}^\dagger$.

There are three other avatars of this models and the analysis made
for the $(2,2)$ model can be done again here.
In particular,  three (not four!) fermionic versions are inequivalent
and we obtain $1+3$  integrable periodic and periodic-twisted 
boundary conditions.

\subsection{The $su(4)$-XXZ model}

This model, which is not an XXC one, 
is based on the fundamental representation
of $su(4)$.  In the notation of \cite{anp} where the XXC models were
further generalized, one has the  $(1,1,1,1;4,4)$ system.
We could  fermionize the $su(3)$ XXZ as was done
for $(1,2)$-XXC model.
We   remain however at four states per site and 
study  the $su(4)$ model.

The Hamiltonian density is given by:
\begin{eqnarray}
\CH_{mm+1}&=& a^\dagger_{1m}a_{1m+1}\left[ x_{23}\,\nbar_{2m}\nbar_{2m+1}
-x^{-1}_{14}n_{2m}n_{2m+1}\right] \\ 
          &+&a^\dagger_{1m+1}a_{1m}\left[ x^{-1}_{23}\nbar_{2m}
\nbar_{2m+1}-x_{14}\,n_{2m}n_{2m+1}\right]  \nonumber\\
          &+&a^\dagger_{2m}a_{2m+1}\left[ x_{24}\,\nbar_{1m}
\nbar_{1m+1}-x^{-1}_{13}n_{1m}n_{1m+1}\right]  \nonumber\\
          &+&a^\dagger_{2m+1}a_{2m}\left[ x^{-1}_{24}\nbar_{1m}
\nbar_{1m+1}-x_{13}\,n_{1m}n_{1m+1}\right] \nonumber\\
          &+&x^{-1}_{12}a^\dagger_{1m}a_{1m+1}a^\dagger_{2m}a_{2m+1}
          +x_{12}\,a^\dagger_{1m+1}a_{1m}a^\dagger_{2m+1}a_{2m} \nonumber\\
          &-&x^{-1}_{34}a^\dagger_{1m}a_{1m+1}a^\dagger_{2m+1}a_{2m}
          -x_{34}\,a^\dagger_{1m+1}a_{1m}a^\dagger_{2m}a_{2m+1} \nonumber\\
&+& \cos\gamma\, P^{(2)}_{mm+1} + i\,\sin\gamma\,(P^{(+)}_{mm+1}
-P^{(-)}_{mm+1})\nonumber
\end{eqnarray}
where 
\bea
P^{(+)}_{mm+1}&=& P^{(-)}_{m+1m}=
n_{1m}n_{2m} (\nbar_{1m+1}\nbar_{2m+1} +n_{1m+1}\nbar_{2m+1}
+ \nbar_{1m+1}n_{2m+1})\\
&+& \nbar_{1m}\nbar_{2m} (n_{1m+1}\nbar_{2m+1}
+\nbar_{1m+1}n_{2m+1}) + n_{1m}\nbar_{2m}\nbar_{1m+1}n_{2m+1}\nonumber\\
P^{(2)}_{mm+1}&=&n_{1m}n_{2m}n_{1m+1}n_{2m+1}+\nbar_{1m}\nbar_{2m}
\nbar_{1m+1}\nbar_{2m+1}\\
&+&n_{1m}\nbar_{2m}n_{1m+1}\nbar_{2m+1}
+\nbar_{1m}n_{2m}\nbar_{1m+1}n_{2m+1}\nonumber
\eea
Let  $a'=\sin(\gamma - \lambda)$, 
$b'=\sin\lambda$, $c'=e^{i\lambda}\sin\gamma$ 
and $d'=e^{-i\lambda}\sin\gamma$.
We find for the Lax  matrix:
\begin{equation}
\CL_m=   \left(\begin{array}{cccc}
            \CL_{11}&-d' a_{1m}a_{2m}&-i\,d'a_{2m}n_{1m}&-i\,d' a_{1m}n_{2m}\\
            -c' a^\dagger_{1m}a^\dagger_{2m} & \CL_{22}&-i\,d' a^\dagger_{1m}\bar{n}_{2m}&i\,d'a^\dagger_{2m}\bar{n}_{1m}\\
            c'a^\dagger_{2m}n_{1m}&-c'a_{1m}\bar{n}_{2m}&\CL_{33}&-i\,d' a^\dagger_{2m}a_{1m}\\
            c'a^\dagger_{1m}n_{2m}&c'a_{2m}\bar{n}_{1m}&-i\,c'a^\dagger_{1m}a_{2m}&\CL_{44}
       \end{array} \right)
 \end{equation}
\beas
&\CL_{11}& = a'n_{1m}n_{2m}
    -b'\left\{ x_{12}\,\nbar_{1m}\nbar_{2m}+i\,x_{13}\,n_{1m}\nbar_{2m}
+i\,x_{14}\,\nbar_{1m}n_{2m}\right\} \\ 
&\CL_{22}& = -a'\nbar_{1m}\nbar_{2m}
    +b'\left\{ x^{-1}_{12}n_{1m}n_{2m}-i\,x_{23}\,n_{1m}\nbar_{2m}
-i\,x_{24}\,\nbar_{1m}n_{2m}\right\} \\ 
&\CL_{33}& = i\,a' n_{1m}\nbar_{2m}
    +b'\left\{ x^{-1}_{13}n_{1m}n_{2m}-x^{-1}_{23}\nbar_{1m}\nbar_{2m}
+i\,x_{34}\,\nbar_{1m}n_{2m}\right\} \\ 
&\CL_{44}& = i\,a'\nbar_{1m}n_{2m}
+b'\left\{ x^{-1}_{14}n_{1m}n_{2m}-x^{-1}_{24}\nbar_{1m}\nbar_{2m}
+i\,x^{-1}_{34}n_{1m}\nbar_{2m}\right\} 
\eeas
The grading, $V$ and $F$ matrices are unchanged.  
As the bosonic model has only diagonal symmetries, 
these survive at the fermionic level and there are no other symmetries,
whatever the values of the  parameters $x$.
The symmetries are the $X^{ii}$ of (\ref{syme}), and they generate
$u(1)\times u(1) \times u(1)$.

\subsection{General remarks}\label{genrem}

Let us pause and recapitulate what we have learned so far. 
Several conclusions arrived at for the foregoing examples are
of a general nature. The first concerns boundary conditions. 
Going back to bosonic variables generates many integrable
systems. The non-local boundary terms are generically of the 
type $E^{a\be}_1 (U\gamma U)_2\cdots (U\gamma U)_{N-1} E^{\be a}_N$, 
with their  hermitian
conjugate.  Thus, to each XXC
system is associated a number of integrable boundary conditions
equal to $1$, for the usual periodic bosonic conditions, 
plus a  number  less than or equal to the number of versions the XXC model has.
This generalizes to the models of \cite{anp}.
The non-locality of the  generalized Jordan-Wigner transformation appears as
a strength rather than  a handicap.

Another important result concerns the symmetries.
We have seen that  symmetry generators of the bosonic
model, which are local in fermionic variables, are also 
symmetries of the fermionic system with periodic fermionic
boundary conditions.  The generators which become non-local
are not symmetries of the fermionic model. However we can consider
their `localized' versions, obtained by a simple deletion of the
non-local string stemming for the fermionization.
These {\it odd} parity operators are  symmetries 
provided the twist parameters are
chosen in a certain way. This also generalizes to the models of \cite{anp}.

All  trigonometric models admit a rational limit. 
The rational limit of $L$-$R$ is obtained by letting $\la\rightarrow\ga\la$,
dividing by $\sin\ga$ and taking the limit $\ga\rightarrow 0$. These
manipulations conserve all the properties of the $L,R$-matrices and can be 
applied directly to the fermionic form, without any other change.
However the eventual extended symmetries which some bosonic models
may enjoy breaks down for their fermionic versions.  

The general form of $v_m$ is obvious:
\beq
v_m= \exp\left[\frac{i\pi}{2} \sum_{k=1}^{m-1} \sum_{j=1}^d 
(n_{jk}-1)\right]
\eeq
The four phases $\pm 1$ and  $\pm i$ arise from the braiding
relations between $v_m$ and the other fermionic operators. The particular structure of $V_m$ is fixed by the choice of representation
of $\CC_{2d}$. This is true also for the grading and the matrix $F$.
The matrix elements of the diagonal matrix  $F$ are found by 
solving a redundant system of very simple linear equations.
It is in fact enough to solve part of this system, although
we have solved it completely for the systems studied here.

Most  of these general results extend straightforwardly  to the generalized 
Hubbard models considered in the following section.

\section{Hubbard models}

The  Jordan-Wigner transformation turns the  Hubbard model into 
a spin-chain of two coupled XX chains \cite{sh12}. In \cite {zm12,ff}
the  construction of bosonic Hubbard models was generalized to 
all XX models. The latter are XXC models at their `free-fermions'
point: $\ga=\frac{\pi}{2}$.
The fermionic integrable structure of the original Hubbard model
was obtained  by generalizing the method used for the
XXZ model. We first briefly recall this method 
\cite{woa} and then generalize it to fermionize  generalized
Hubbard models by fusing  some of the models obtained in the 
preceding section. The left and right copies are labeled $\uparrow$
and $\downarrow$, respectively. 

\subsection{The Hubbard model}\label{huborig} 

The Hubbard Hamiltonian  is that of a spin-$\frac{1}{2}$ electron
hopping on a lattice with an on-site Coulomb interaction.
There are four possible states per site and:
\beq
\CH_2=-\sum_m \sum_{s=\uparrow,\downarrow}
\left(x_s\,a^\dagger_{sm+1} a_{sm}  + 
x_s^{-1}a^\dagger_{sm} a_{sm+1}\right)
+ U \sum_m (2n_{\uparrow m}-1)(2n_{\downarrow m}-1)
\eeq
Its bosonic version \cite{sh12} is given by
\beq
H_2= \sum_m \left(x^{-1}_{\uparrow}\sig^+_m \sig^-_{m+1} 
+x_{\uparrow} \,\sig^-_m \sig^+_{m+1}
+x_{\downarrow}^{-1}\tau^+_m \tau^-_{m+1} 
+x_{\downarrow}\, \tau^-_m \tau^+_{m+1}\right)
+U\sum_m \sig^z_m \tau^z_m
\eeq
where $\sig^i$ and $\tau^i$ are two commuting copies of Pauli matrices. 
Note that this model is integrable without any  constraint 
on the parameters $x_s$.  Its $L$-matrix is given by
\beq
L(\la)=I(h)\, L_{\uparrow}(\la)\otimes L_{\downarrow}(\la)\, I(h)
\eeq
where
\beq
h=h(\la)=\frac{1}{2}{\rm Arcsinh}[ U\sin(2\la)] \quad{\rm and} 
\quad  I(h) =\exp(\frac{h}{2}\sig^z_{\uparrow 0}\tau^z_{\downarrow 0})
\eeq
The relation between  $h$ and $\la$ is the same for all models
of this section. The $V$-matrix
is defined as follows:
\beq
V_m=V_{\uparrow m}\otimes V_{\downarrow m}
\eeq
where 
\beq
V_{\uparrow m} =  \left(\begin{array}{cc}
            v_{\uparrow m} & 0  \\
            0 & v_{\uparrow m}^{-1}
       \end{array} \right)\qquad {\rm and}\qquad
V_{\downarrow m} =  \left(\begin{array}{cc}
             v_{\uparrow N+1} v_{\downarrow m} & 0  \\
            0 & v_{\uparrow N+1}^{-1} v_{\downarrow m}^{-1}
       \end{array} \right)
\eeq
and $v_{sm}=\exp\left(\frac{i\pi}{2} \sum_{j=1}^{m-1} (n_{sj}-1)\right)$,
$s=\uparrow,\downarrow$.
The factors $I(h)$ are diagonal and act in the auxiliary spaces,
and are therefore not fermionized. The fermionized $\CL$-matrix
reads
\bea
\CL_m &=& I(h)\, \tilde{\CL}_{\uparrow m}\ots {\CL}_{\downarrow m}\,  I(h)
\label{cou1}\\
\tilde{\CL}_{\uparrow m}&=& S\,\CL_{\uparrow m}\, S^{-1} \quad ,
\quad S=\left(\begin{array}{cc}
            1 & 0 \\
            0 & i 
       \end{array} \right)\label{cou2}
\eea
with a grading defined by $P(1)=0$ and $P(2)=1$. 
Both $\CL_{\uparrow m}$ and  $\CL_{\downarrow m}$ matrices
are copies of matrix (\ref{clxxz}) where $\ga=\pi/2$.
Note that $S$ is determined up to an overall factor;
we take $S_{11}=1$.

The $\CR$-matrix is also obtained through a similarity transformation,
\beq
\check{\CR}(\la_1,\la_2)= F^{-1} \check{R}(\la_1,\la_2)\, F
\eeq  
where
\beq
F={\rm diag}(1,1,i,i,i,i,1,1,-1,-1,-i,-i,-i,-i,-1,-1)
=\sig^z\otimes 
\left(\begin{array}{cccc}
            1 & 0 & 0 & 0 \\
            0 & i & 0 & 0 \\
            0 & 0 & i & 0 \\
            0 & 0 & 0 & 1
       \end{array} \right)\otimes \bI_2
\eeq
The corresponding grading for relation (\ref{frllc}) is
given by: $P(1)=P(4)=0$ and $P(2)=P(3)=1$.
Two spectral parameters appear instead of just a difference
because the $R$-matrix of the Hubbard model is not additive;
this  is an unusual feature of {\it all\/} the Hubbard models considered here.

The integrable hierarchy defined by the Hubbard Hamiltonian is known to have
an $SO(4)$ group symmetry \cite{hl,yang,yazh,pern,affl,gomus}. 
The two $u(1)$ generators are  inherited
from the two fermionic constituents. For  $x_\uparrow=x_\downarrow=\pm 1$,
the symmetry further extends and the generators of the Lie algebra
$so(4)=su(2)\times su(2)$ are given by:
\bea
&S^+=\sum_m a^\dagger_{\uparrow m} a_{\downarrow m}\;,\;\;\;
S^-=\sum_m a^\dagger_{\downarrow m} a_{\uparrow m}\;,\;\;\;
S^z=\frac{1}{2}\sum_m (n_{\uparrow m}-n_{\downarrow m})&\\
&\eta^+=\sum_m (-1)^m a^\dagger_{\uparrow m} a^\dagger_{\downarrow m}\;,\;\;\;
\eta^-=\sum_m (-1)^m a_{\downarrow m} a_{\uparrow m}\;,\;\;\;
\eta^z=\frac{1}{2}\sum_m (n_{\uparrow m}+n_{\downarrow m}-1)&
\eea
This extended symmetry arises {\it after} coupling of the two XX 
models. It exponentiates to the $SO(4)$ group 
symmetry and  characterizes many  physical features 
of the one-dimensional Hubbard model (see for instance \cite{ke}).

The fermionization of the bosonic Hubbard models of \cite{ff}
is carried out in a similar manner.
We give below the results for the fusion of some of the models
studied in the preceding section. 

\subsection{The $(1,1)\times(2,2)$ model}\label{h1122}

We use as right copy  the first version of the $(2,2)$-XXC model.
The resulting model has a local space of dimension eight.
The Hamiltonian reads:
\beq
\CH_2 = \CH_2^\uparrow + \CH_2^\downarrow + U \sum_m (2n_{\uparrow m}-1)
(2n_{\downarrow 1 m}-1)(2n_{\downarrow 2 m}-1)\label{hamil}
\eeq
where $\CH_2^\uparrow$ corresponds to (\ref{fh11})  and 
$\CH_2^\downarrow$ to (\ref{fh22}),
both at $\gamma=\pi/2$.   
The $V$-matrix is a tensor product of $V_{\uparrow m}$ of the preceding
section on the left,  and of 
\bea
V_{\downarrow m} &=&\left(\begin{array}{cccc}
v_{\uparrow N+1} v_{\downarrow m}&0&0&0\\
0&v_{\uparrow N+1} v_{\downarrow m}&0&0\\
0&0&v_{\uparrow N+1}^{-1} v_{\downarrow m}^{-1}&0\\
0&0&0&v_{\uparrow N+1}^{-1} v_{\downarrow m}^{-1}
\end{array}\right)\\ 
& & \\
v_{\downarrow m}&=& \exp\left(\frac{i\pi}{2}
\sum_{i=1}^{m-1}\sum_{j=1}^2 (n_{\downarrow j i}-1)\right)
\eea
on the right.
This yields for the Lax matrix $\CL$:
\beq
\CL_m = I(h)\, \tilde{\CL}_{\uparrow m}\ots {\CL}_{\downarrow m}\,  I(h)
\eeq
where 
\beq
I(h)= \exp\left(\frac{h}{2} \sig^z_{\uparrow 0}\gamma^5_{\downarrow 0}\right)
\quad {\rm and} \quad \gamma^5={\rm diag}(1,1,-1,-1)
\eeq
$\tilde{\CL}_{\uparrow m}$ has been defined in the preceding section
while $\CL_{\downarrow m}$ is a copy of (\ref{l22xxc}).
The grading above is:
\beq
P_{\uparrow}(1)=0\;,\;\; P_{\uparrow}(2)=1\;,\;\;
P_{\downarrow}(1)=P_{\downarrow}(2)=0\;,\;\;
P_{\downarrow}(3)=P_{\downarrow}(4)=1
\eeq
The  $64\times 64$  matrix $F$    is given by 
\beq
F=\sig^z \otimes {\rm diag}(1,i,-1,-i,i,1,i,1)\otimes \bI_4
\eeq
and the grading here is:
\beq
P(1)=P(2)=P(7)=P(8)=0 \quad ,\quad \quad P(3)=P(4)=P(5)=P(6)=1
\eeq
This grading is used in the supertrace.

For all $x$'s on the right equal to each other,
the symmetry of this Hubbard model is 
$u(1)_\uparrow \times su(2)_\downarrow \times su(2)_\downarrow \times u(1)_\downarrow$. 
Note that is is possible to take the second (or third)
copy of the $(2,2)$ model. The Hamiltonian (\ref{hamil})
is modified accordingly. By choosing the $x$'s on the right as indicated
in section (\ref{avatar2}), one restores the symmetry to:
$u(1)_\uparrow \times gl(1|1)_\downarrow \times gl(1|1)_\downarrow$.
It is also possible to choose the model $(1,3)$ instead of $(2,2)$, and 
the appropriate choice of $x$'s, with a 
resulting symmetry of: 
$u(1)_\uparrow \times gl(2|1)_\downarrow \times u(1)_\downarrow$.
The $S$, $V$, $F$ matrices and the gradings are unchanged.

\subsection{The $(2,2)\times(2,2)$ model}

Again we use two copies of the  first version of the $(2,2)$-XXC model.
The dimension of the local Hilbert space is sixteen.
The Hamiltonian is now given by:
\beq
\CH_2 = \CH_2^\uparrow + \CH_2^\downarrow + U \sum_m (2n_{\uparrow 1 m}-1)
(2n_{\uparrow 2 m}-1)(2n_{\downarrow 1 m}-1)(2n_{\downarrow 2 m}-1)
\eeq
where $\CH_2^\uparrow$  and $\CH_2^\downarrow$ corresponds to two copies
of (\ref{fh22})  at $\gamma=\pi/2$.  
The numerical coupling matrix is given by: $I(h)=\exp(\frac{h}{2} 
C_{\uparrow 0}C_{\downarrow 0})$ where 
$C=\ga^5={\rm diag}(1,1,-1,-1)$.
The $V$-matrices are: 
\beq
\begin{array}{l}
v_{s m}= \exp\left(\frac{i\pi}{2}
\sum_{i=1}^{m-1}\sum_{j=1}^2 (n_{s j i}-1)\right)\;,
\quad s=\uparrow,\downarrow\\ 
\\
V_m ={\rm diag}(v_{\uparrow m},v_{\uparrow m},v_{\uparrow m}^{-1},
v_{\uparrow m}^{-1})\otimes {\rm diag}(v_{\uparrow N+1} v_{\downarrow m},
v_{\uparrow N+1} v_{\downarrow m}, v_{\uparrow N+1}^{-1} v_{\downarrow m}^{-1},
v_{\uparrow N+1}^{-1} v_{\downarrow m}^{-1})
\end{array}
\eeq
The  $\CL_{sm}$ matrices are copies of 
matrix (\ref{l22xxc}) at $\ga=\pi/2$.
The grading in (\ref{cou1}--\ref{cou2})  is given by: $P(1)=P(2)=0$, $P(3)=P(4)=1$, and  $S={\rm diag}(1,-1,i,i)$ for the `up' copy.  
For  (\ref{frllc}) and the supertrace the grading is:
\bea
\begin{array}{l}
P(1)=P(2)=P(5)=P(6)=P(11)=P(12)=P(15)=P(16)=0 \\
P(3)=P(4)=P(7)=P(8)=P(9)=P(10)=P(13)=P(14)=1
\end{array}
\eea
The  $256\times 256$  matrix $F$    is given by 
\beq
F=\sig^z\otimes \bI_2\otimes {\rm diag}
(1,-1,i,i,-1,1,-i,-i,i,-i,1,1,i,-i,1,1)\otimes\bI_4
\eeq

For all $x$'s, on both left and  right independently, equal to each other,
the symmetry of this Hubbard model is 
$su(2)_\uparrow \times su(2)_\uparrow\times u(1)_\uparrow 
\times su(2)_\downarrow \times su(2)_\downarrow \times u(1)_\downarrow$. 
As indicated at the end of section (\ref{h1122}),
it is possible to fuse any two copies of  the other versions of
$(2,2)$ , or $(1,3)$. The resulting symmetry changes accordingly,
with the appropriate choices of  $x$'s.
The gradings and $V$, $S$, $F$ are unchanged.
It is possible to couple also $(1,2)$ models, or in general any two
XX models in their fermionic form. A fermionic Hubbard Hamiltonian
is simply:
\beq
\CH_2=\CH_2^\uparrow +\CH_2^\downarrow +U\sum_m \CC_m^\uparrow \CC_m^\downarrow
\eeq

\subsection{Symmetry and Integrability}\label{symii} 

There are many models of itinerant electrons  in the literature which qualify 
as such for the name Hubbard model. 
However, most of these models have an additive 
$R$-matrix, or do not share the same algebraic structure of 
the original Hubbard model (the Bariev model). Here we chose to use   
the term Hubbard models for a class of models
which share the same integrable and algebraic structure as the original model
along with a similar structure for the 
non-additive $R$-matrix. These bosonic multistates Hubbard models were
constructed and studied in \cite{zm12,ff}.
(Lax pairs were derived in \cite{ys}.) They correspond
to {\it arbitrary} up and down copies of the generalized XX models,
with and  on-site coupling. This  structure, which is the natural 
generalization of the Hubbard model one,  has been fermionized
here,  and we obtained
the first multispecies generalizations
written in fermionic form.

Recall that (see section (\ref{huborig}))
the Hubbard model has,  after coupling of its two components,
an extended symmetry not shared by its two independent components.
This symmetry also exponentiates to a group symmetry. 
It is therefore natural to ask whether the new models we obtained here
have such a property. We have looked for additional symmetries
of the new Hubbard models, and found  it is
unlikely that they exist. This negative result can be explained
by a  conflict between higher symmetries and the
strictures of integrability. The form of the fermionized XXC models 
shows that they are not  constructed out of group invariants
and their symmetries do not exponentiate.
Coupling does not solve this problem. There is however
a finite number of {\it discrete} symmetries
which appear; they correspond to fermion species interchanges.   
In fact one could easily build models of itinerant electrons
with larger symmetries. However integrability breaks down the symmetry
group manifold to a discrete set of points, some corresponding to
the  symmetries of the XXC blocks and some appearing only for the
Hubbard models.

\section{Non integrable models}

The fermionized Hamiltonians seen so far were obtained directly from
their bosonic counterparts without use of any integrability criterion.
The generalized fermionization procedure is independent of any integrable
structure. We now fermionize two non integrable models of 
considerable physical interest. 

\subsection{The spin-$1$ XXZ Heisenberg  chain}

The bosonic Hamiltonian is given by:
\beq
H_2=\sum_m \left( S^x_m S^x_{m+1} +S^y_m S^y_{m+1} + \Delta_z\,
S^z_m S^z_{m+1}\right)\label{hheisen}
\eeq
where the $S^i$ form a spin one representation of $su(2)$, at every site. 
For arbitrary $\Delta_z$ the diagonal operator $S^z=\sum_m S^z_m$ commutes 
with $H_2$. For  $\Delta_z=1$, $S^{x,y}=\sum_m S^{x,y}_m$ in addition 
commute with the Hamiltonian. 

To fermionize this model one  has  four choices of embeddings.
We  chose the same one as in section (\ref{3dim}). 
All states  with double occupancy on at least   one site are therefore 
excluded.
The fermionization then yields:
\begin{eqnarray}
\CH_{mm+1}&=& (a^\dagger_{1m+1}a_{1m}  + a^\dagger_{1m}a_{1m+1})\,\nbar_{2m}\nbar_{2m+1} \label{xxzs1}\\
&+&a^\dagger_{1m+1}a^\dagger_{2m}a_{1m}a_{2m+1} + a^\dagger_{1m}a^\dagger_{2m+1}a_{1m+1}a_{2m}\nonumber\\
&+&\pi_{m-1} \,(a^\dagger_{2m+1}a_{1m}a_{1m+1}
+a^\dagger_{1m+1}a^\dagger_{1m}a_{2m+1})\,\nbar_{2m}\nonumber\\ 
&+&\pi_{m-1}\, (a^\dagger_{1m+1}a^\dagger_{1m}a_{2m} 
+a^\dagger_{2m}a_{1m}a_{1m+1})\,\nbar_{2m+1} \nonumber\\ 
&+& \Delta_z \,\nbar_{1m}\nbar_{1m+1} (2n_{2m}-1)(2n_{2m+1}-1)\nonumber
\end{eqnarray}
where
\beq
\pi_0\equiv 1 \quad , \quad \pi_{m-1} = (2n_{11}-1)(2n_{21}-1)\cdots\cdots
(2n_{1m-1}-1)(2n_{2m-1}-1)
\eeq
Note that, contrary to all the integrable models we have  studied so far,
there are non-local contributions associated with odd-parity combinations
in the Hamiltonian density. It is also easily seen that
its action is stable  on states without double occupancy.
Let $p$ be the projection operator (\ref{proj}) 
defined in section (\ref{3dim}). 
The action of $\CH_2=\sum_m \CH_{mm+1}$ on the $3^N$-dimensional subspace
is equivalent to the action of $p \CH_2 p$ on the $4^N$-dimensional space. 

We may remove the  $\pi_{m-1}$'s, leaving their factors in,
wherever they appear in  (\ref{xxzs1}), and obtain
a local density which combines bosonic and fermionic pieces. 
The local spin-one fermionic version the XXZ Heisenberg
model may then be defined this way. The diagonal operator
$X^z=\sum_m \nbar_{1m}(2n_{2m}-1)$ still commutes with the Hamiltonian
with periodic boundary conditions. 
However as seen for the spin-$\frac{1}{2}$ Heisenberg chain,
the extended symmetry at $\Delta_z=1$ does not survive the fermionization.

\subsection{The spin-$\frac{3}{2}$  XXZ Heisenberg chain}

The bosonic Hamiltonian is given by (\ref{hheisen}) where
now the $S^i$ span a spin $\frac{3}{2}$ representation of $su(2)$.
Here the Hilbert space of the chain is $4^N$-dimensional.
We find for the fermionized Hamiltonian density:
\begin{eqnarray}
\CH_{mm+1}&=&\frac{1}{2}\, a^\dagger_{1m}a_{1m+1}\left[ 3\,a^\dagger_{2m}a_{2m+1}+3\,a^\dagger_{2m}a^\dagger_{2m+1}+3\,a_{2m}a_{2m+1}
\right.\\ 
&+&\left. 3\,a_{2m}a^\dagger_{2m+1}+4\,
\nbar_{2m}\nbar_{2m+1}\right]\nonumber \\ 
     &+&\frac{1}{2}\, a^\dagger_{1m+1}a_{1m}\left[ 3\,a^\dagger_{2m+1}a_{2m}+3\,a^\dagger_{2m+1}a^\dagger_{2m}+3\,a_{2m+1}a_{2m}
\right. \nonumber\\
     &+&\left. 3\,a_{2m+1}a^\dagger_{2m}+4\,\nbar_{2m}\nbar_{2m+1}\right] 
 \nonumber\\
     &+&\sqrt{3}\,a^\dagger_{1m}a^\dagger_{1m+1}
            \left[ a^\dagger_{2m}\nbar_{2m+1}-a_{2m}\nbar_{2m+1}
                 -a^\dagger_{2m+1}\nbar_{2m}-a_{2m+1}\nbar_{2m} 
\right] \,\pi_{m-1}\nonumber\\
     &+&\sqrt{3}\,a_{1m}a_{1m+1}
\left[ a^\dagger_{2m}\nbar_{2m+1}-a_{2m}\nbar_{2m+1}
+a^\dagger_{2m+1}\nbar_{2m}+a_{2m+1}\nbar_{2m} \right]\, \pi_{m-1}
\nonumber\\
&+&\frac{\Delta_z}{4}\,( 2n_{1m}-1 )( 4n_{2m}-1)( 2n_{1m+1}-1 )( 4n_{2m+1}-1)
\nonumber
\end{eqnarray}
Remarks similar to those  made for the spin-$1$ case hold here. 
In particular a local density may be defined by the deletion
of $\pi_{m-1}$.
The diagonal commuting operator is given by
\beq
X^z=\frac{1}{2}\sum_m  ( 2n_{1m}-1 )( 4n_{2m}-1)
\eeq

\section{Conclusion}

We have introduced a generalized Jordan-Wigner transformation 
which allows the fermionization/bosonization of integrable and 
non-integrable spin chains. Several variants
of this transformation were described. 
We have shown on various examples how the
mapping works. We also obtained the first
fermionic versions of generalized Hubbard models. 

The study of  particular models  allowed to infer some general results.
In particular we showed that the non-conservation of boundary conditions,
far from being a `shortcoming', is in fact an advantage and 
generates many new integrable boundary
conditions of the periodic-twisted type.
Another issue is the mapping of  certain local  operators to non-local 
ones.  However we have shown that,  the non-local fermionic
operators can be made local and the  symmetry of the model turned 
into supersymmetry, without any loss of integrability. 
Fermionic realizations of Lie algebras and superalgebras
appeared naturally in this context.  

Periodic boundary conditions were considered here. But the fermionization
method can be directly applied to models with other types of boundary
conditions, within or without the framework of integrability.

Fermionization and bosonization of one-dimensional systems are
an expression of a local equivalence between a bosonic
language and a fermionic one. This is clearly seen at the Hamiltonian
level, for integrable and non integrable models. The fermionization
of the $L$-matrix is a local procedure. The important point is to notice
that a periodic boundary condition breaks this equivalence,
while an open one, where  $H_{N1}$ is dropped, does not. (Integrability 
may be lost but this is a secondary point.) 
The issue of boundary conditions is intrinsic to any 
fermionization scheme.

The present work was near completion   
when \cite{gomu} appeared. The authors of \cite{gomu}
proposed a fermionization scheme which applies {\it only} 
to integrable models. 
Their main example corresponds to the one in section (\ref{3dim}), at 
$\gamma=\pi/2$. The Hamiltonians are given by the same expressions while
$\CL$-matrices have the same structure. The phases $\pm i$ do not 
appear, and this is general, in the method of \cite{gomu}. 
It would be interesting and instructive to find, for integrable systems, 
the exact relation between the two methods. 
However let us stress that boundary issues invariably arise for any fermionization method, and  only the foregoing approach tackles this issue.
The  method of G\"ohmann and Murakami and the generalized 
Jordan-Wigner transformation are therefore complementary.

A given one-dimensional Hamiltonian density can be transposed on a 
higher-dimensional lattice. 
While in one-dimension bosons and fermions are closely related,
the relationship breaks down in higher dimensions.
This is one important motivation for finding the fermionic expression 
of a bosonic model in one-dimension, before going to higher dimensions.
The physical properties of a model
may change drastically with dimension. 
As noted in section (\ref{symii}),
the study of one-dimensional models would shed some light 
on the interplay between integrability and symmetries. 
But it is only on a two-dimensional square lattice with nearest-neighbor
interaction that
the Hubbard model  exhibits $d$-wave 
superconductivity. It would be worthwhile
to study the new Hubbard models in one and two dimensions.

\bigskip\ {\bf Acknowledgement:} Z.M.   would like to thank N.E. Bickers
for sharing some of his insight on the Hubbard model.


\begin{thebibliography}{30}

\bibitem{fms} {\it Conformal Field Theory},
Ph. Di Francesco, P. Mathieu and D. S\'en\'echal, Springer (1997).
\bibitem{jowi} P. Jordan and E. Wigner, Z. Phys. {\bf 47} (1928) 631.
\bibitem{puzh} F.-C. Pu and B.-H. Zhao, Phys. Lett. A {\bf 118} (1986) 77--81.
\bibitem{sh12} B.S. Shastry, Phys. Rev. Lett. {\bf 56}, (1986) 1529--1531; 
Phys. Rev. Lett. {\bf 56}, (1986) 2453--2455.
\bibitem{bariev} R.Z. Bariev, J. Phys. {\bf A 24} (1991) L549--L553,
J. Phys. {\bf A 24} (1991) L919--L923.
\bibitem{zm12} Z.~Maassarani, Phys. Lett. A {\bf 239} (1998) 187--190; 
Mod. Phys. Lett. B {\bf 12} (1998) 51--56.
\bibitem{ff}  Z.~Maassarani, {\it Hubbard Models as Fusion Products
of Free Fermions},  Int. J. Mod. Phys. B {\bf 12} (1998), in press,
LAVAL-PHY-26/97, cond-mat/9711142.
\bibitem{gomu} F. G\"ohmann and S. Murakami,
{\it Fermionic representation of integrable lattice systems}, cond-mat/9805129.
\bibitem{zinn} {\it Quantum Field Theory and Critical Phenomena}, 
J. Zinn-Justin, Oxford (1990). 
\bibitem{xxc} Z.~Maassarani, Phys. Lett. A {\bf 244} (1998) 160--164. 
\bibitem{woa}  E. Olmedilla,  M. Wadati and Y. Akutsu, 
J. Phys. Soc. Jpn. {\bf 56}, (1987) 2298--2308.
\bibitem{qism1} P.P. Kulish and E. K. Sklyanin, in {\it Integrable
Quantum Field Theories}, Tv\"arminne (1981) eds. J. Hietarinta and C. Montonen,
Lect. Notes in Phys. 151, Springer.  
\bibitem{qism2} L.D.~Faddeev in {\it Recent Advances in Field Theory and 
Statistical Mechanics}, Les Houches (1982) eds. J.-B. Zuber and 
R. Stora, North Holland  (1984); and references therein.
\bibitem{kbi} {\it Quantum Inverse Scattering Method and Correlation Functions},
V.E. Korepin, N.M. Bogoliubov and A.G. Izergin, Cambridge University Press
(1993). 
\bibitem{ks} P.P. Kulish and E.K. Sklyanin, J. Sov. Math. {\bf 19} (1982) 1596.
\bibitem{ppk} P.P. Kulish, J. Sov. Math. {\bf 35} (1985) 2648.
\bibitem{aars} F.C. Alcaraz, D. Arnaudon, V. Rittenberg and M. Scheunert,
Int. J. Mod. Phys. A {\bf 9} (1994) 3473--3496; and references therein.
\bibitem{mm}  Z.~Maassarani and P.~Mathieu, 
Nucl. Phys. B {\bf 517}, Nos. 1--3 (1998) 395--408.
\bibitem{anp} Z. Maassarani, {\it Multiplicity $A_m$ models}, LAVAL-PHY-20/98, 
solv-int/9805009.
\bibitem{hl} O.J. Heilmann and E.H. Lieb, Ann. N.Y. Acad. Sci. {\bf 172}
(1971) 584. 
\bibitem{yang} C.N. Yang, Phys. Rev. Lett. {\bf 63} (1989) 2144.
\bibitem{yazh} C.N. Yang and S.C. Zhang, Mod.  Phys.  Lett. B {\bf 4} (1990)
759.
\bibitem{pern} M. Pernici, Europhys. Lett. {\bf 12} (1990) 75.
\bibitem{affl} I. Affleck, in {\it Physics, Geometry and Topology},
ed. H.C. Lee, Plenum Press (New York, 1990). 
\bibitem{gomus} F. G\"ohmann and S. Murakami, J. Phys. A {\bf 30} (1997) 5269;
M. Shiroishi, H. Ujino and M. Wadati, J. Phys. A {\bf 31}
(1998) 2341--2358.
\bibitem{ke}  {\it Exactly solvable models
of strongly correlated electrons}, V.E. Korepin and F.H.L. E\ss ler, 
World Scientific (Singapore). 
\bibitem{ys} R. Yue and R. Sasaki, {\it Lax pair for $SU(n)$ Hubbard model},
YITP-98-5, cond-mat/9801193. 


\end{thebibliography}
\end{document}